\documentclass[journal]{IEEEtran}
\usepackage{graphicx}
\usepackage{amssymb}
\usepackage{amsmath}
\usepackage{makecell}
\usepackage{multicol}
\usepackage{authblk}
\usepackage{float}
\usepackage[numbers]{natbib}
\usepackage{xcolor}
\begin{document}

\title{Stochastic Coded Offloading Scheme for Unmanned Aerial Vehicle-Assisted Edge Computing}

\author{Wei Chong Ng, Wei Yang Bryan Lim,
Zehui Xiong, Dusit Niyato, \emph{IEEE Fellow}, Chunyan Miao,\\ Zhu Han, \emph{IEEE Fellow}, and Dong In Kim, \emph{IEEE Fellow}
\thanks{WC. Ng and WYB. Lim are with Alibaba Group and Alibaba-NTU Joint
Research Institute, Nanyang Technological University, Singapore}% <-this % stops a space%
\thanks{ZH. Xiong is with Singapore University of Technology and Design, Singapore}
\thanks{D. Niyato is with School of Computer Science and Engineering, Nanyang
Technological University, Singapore.}
\thanks{C. Miao is with Joint NTU-UBC Research Centre of Excellence in
Active Living for the Elderly (LILY) and School of Computer Science and
Engineering, Nanyang Technological University, Singapore}
\thanks{Z. Han is with the Department of Electrical and Computer Engineering in the University of Houston, Houston, TX 77004 USA, and also with the Department of Computer Science and Engineering, Kyung Hee University, Seoul, South Korea, 446-701.}
\thanks{D. I. Kim is with the Department of Electrical and Computer Engineering, Sungkyunkwan University (SKKU), Suwon 16419, South Korea}}

\maketitle

\begin{abstract}
Unmanned aerial vehicles (UAVs) have gained wide research interests due to their technological advancement and high mobility. The UAVs are equipped with increasingly advanced capabilities to run computationally intensive applications enabled by machine learning techniques. However, because of both energy and computation constraints, the UAVs face issues hovering in the sky while performing computation due to weather uncertainty. To overcome the computation constraints, the UAVs can partially or fully offload their computation tasks to the edge servers. In ordinary computation offloading operations, the UAVs can retrieve the result from the returned output. Nevertheless, if the UAVs are unable to retrieve the entire result from the edge servers, i.e., straggling edge servers, this operation will fail. In this paper, we propose a coded distributed computing approach for computation offloading to mitigate straggling edge servers. The UAVs can retrieve the returned result when the number of returned copies is greater than or equal to the recovery threshold. There is a shortfall if the returned copies are less than the recovery threshold. To minimize the cost of the network, energy consumption by the UAVs, and prevent over and under subscription of the resources, we devise a two-phase Stochastic Coded Offloading Scheme (SCOS). In the first phase, the appropriate UAVs are allocated to the charging stations amid weather uncertainty. In the second phase, we use the $z$-stage Stochastic Integer Programming (SIP) to optimize the number of computation subtasks offloaded and computed locally, while taking into account the computation shortfall and demand uncertainty. By using a real dataset, the simulation results show that our proposed scheme is fully dynamic, and minimizes the cost of the network and UAV energy consumption amid stochastic uncertainties.
\end{abstract}

\begin{IEEEkeywords}
%IEEE, IEEEtran, journal, \LaTeX, paper, template.
Unmanned Aerial Vehicles, Coded Distributed Computing, Stochastic Integer Programming, Task Allocation, Internet-of-Things
\end{IEEEkeywords}

\section{Introduction}
Due to the rapid advancement of Internet of Things (IoT) enabled technologies, the number of wirelessly connected devices is increasing exponentially~\cite{du2017contract} and generating huge amounts of data~\cite{peng2021joint}. There are many new real-time applications enabled by wirelessly connected devices, such as augmented/virtual reality~\cite{von2019increasing} and smart cities~\cite{zanella2014internet} that are delay-sensitive. For example, it is important to know the real-time traffic~\cite{djahel2013adaptive}/parking~\cite{vlahogianni2016real} information to regulate traffic flow. The increase in wirelessly connected devices exerts a tremendous burden on the wireless communication infrastructure. For example, in many urban areas that are covered by dense skyscrapers or when the end-users are in congested regions or at high-speed vehicular network~\cite{7875131}, the content in the static roadside units (RSUs)/base stations (BSs) cannot be successfully delivered to the end-users.

One solution is to deploy Unmanned Aerial Vehicles (UAVs), also known as drones, to act as an airborne BS to collect and process data from the terrestrial nodes~\cite{wang2018network,cheng2018air}. UAVs are in different shapes and sizes, such as fixed wings or multi-rotors, and they can maintain a line-of-sight communication with the end-users to provide a better quality of service (QoS). Furthermore, UAVs can be flexibly deployed to inaccessible terrains or disaster relief operations, e.g., due to their size and mobility. Moreover, wireless connections can be established without a fixed infrastructure to extend communication coverage. However, apart from all those benefits, UAVs are faced with energy constraints~\cite{koulali2016green}, and thus, they cannot complete their computation tasks if the energy utilization is not scheduled correctly.

%Other than commercial use, UAVs can also be used as personal drones for photo-taking; people want the photo to be taken from “impossible angles,” such as few meters away from the ground. The disabled person can also use the UAVs to act as “eyes” (via streaming video) and “ears” (via streaming audio)~\cite{DBLP:journals/corr/Loke15}. 

In this paper, we consider a network contains various UAVs, mobile charging stations, and edge servers that are attached to the BSs to run applications such as traffic monitoring~\cite{ro2007lessons,du2018auction}. The UAVs are required to perform computation, e.g., distributed matrix multiplication, as it is central to many modern computing applications, including machine learning and scientific computing~\cite{dutta2019optimal,liu2020federated} in applications such as post-disaster relief assistance~\cite{tang2019integration} and crowd detection~\cite{motlagh2017uav}. To alleviate some of the battery constraints of the UAVs, the matrix multiplication can be offloaded to ground-based edge servers for processing. The matrix multiplication in the UAVs can be accelerated by scaling the multiplication out across many distributed computing nodes in BSs or edge servers~\cite{yu2017polynomial} known as the workers. However, there is a significant performance bottleneck that is the latency in waiting for the slowest workers, or “stragglers” to finish their tasks~\cite{yu2017polynomial}. Coded distributed computing (CDC) is introduced to deal with stragglers in distributed high-dimensional matrix multiplication. In CDC, the computation strategy for each worker is carefully designed so that the UAV only needs to wait for the fastest subset of workers before recovering the output~\cite{yu2017polynomial}. The minimum number of workers that the UAV has to wait for to recover their results is known as the recovery threshold.

Apart from using the CDC technique to mitigate stragglers, there are three challenges in this network. The first challenge is the weather uncertainty. If the UAV is not properly allocated, it may not withstand the strong wind if its engines are not sufficiently powerful and the battery capacity is small. The second challenge is demand uncertainty. Typically, the edge servers in the BSs require the users to pay a subscription fee in advance, e.g., monthly subscription, so that the users, i.e., UAVs, can use the offloading service. For instance, in matrix multiplication, the size of the matrices, which is the demand, is not always the same. If the actual matrix size is very small, it will be cheaper to perform the local computation within the UAV. Therefore, an uncertainty of actual demand can result in an over-and under-subscription problem. The third challenge is the shortfall uncertainty. Once after the UAVs are allocated, they can perform full local computation, full offload or partial offloading. If the UAV decides to offload the computation to the edge servers in BSs, there is shortfall uncertainty that the copies cannot be returned by any edge servers on time to the UAV due to delays and link failure~\cite{wang2019batch}. It means that the total copies that the UAV has is less than the recovery threshold, where each copy is a sub-portion of matrices involved in the matrix multiplication operation. Therefore, the UAV has to pay a correction cost to re-compute the number of shortfalls locally or re-offload them to match the recovery threshold. This correction cost also involves a hovering cost as the UAVs have to hover in the sky throughout the re-computation. 

To overcome the three challenges mentioned above, we introduce the Stochastic Coded Offloading Scheme (SCOS). SCOS is a two-phase optimization scheme that adopts a CDC technique to reduce the total cost of the network:
\begin{itemize}
    \item \textbf{Phase one (UAV type allocation):} The application owner will first allocate the appropriate UAV to each mobile charging station by considering the weather condition in each time slot. This weather uncertainty is modeled by a two-stage Stochastic Integer Programming (SIP)~\cite{birge2011introduction}.
    \item \textbf{Phase two (task allocation):}
    There are a different number of time-frames/periods within the same time slot. For example, when the morning is the first time slot, each hour is treated as one period, and task allocation occurs in each period. Demand and shortfall are the two uncertainties in task allocation. Instead of performing local computation as the correction action to correct the shortfalls, the same decision options are provided to the UAVs until the $z$ stage. Therefore, $z$-stage SIP is used to model the demand and shortfall uncertainty in various stages.
\end{itemize}
Extensive simulations are performed to evaluate the effectiveness of SCOS. The results show that SCOS can minimize the total cost and the UAVs' energy consumption, especially compared with the traditional deterministic baseline scheme. 

The contributions of this paper are summarized as follows.
\begin{itemize}
    \item The combination/integration yields fully dynamic on-demand computing solutions for emerging applications such as road traffic prediction for autonomous vehicles in which traditional approaches are ineffective due to their rigid and fixed deployment.
    \item Our SCOS is able to provide strategic scenario-based decision that adapts well with the weather condition in which the current solutions for UAVs are limited.
    \item The proposed SCOS can minimize the UAVs' overall costs by optimizing the task allocation. At the same time, it can also minimize all the UAVs' energy consumption. The optimal solution is achieved by considering both the demand and shortfall uncertainty.
    \item From the performance evaluation, we use the real data to validate that SCOS is the optimal scheme when the performance is compared with the Expected-Value Formulation (EVF) and random scheme.
\end{itemize}

The remainder of the paper is organized as follows: In Section~\ref{related_work}, we review the related works. In Section~\ref{system_model}, we present the system model. In Sections~\ref{UAV allocation} and~\ref{Task allocation} we formulate the problem. We discuss and analyze the simulation result in Section~\ref{simulation}. Section~\ref{conclusion} concludes the paper.

\section{Related Work}\label{related_work}
\subsection{UAV-enabled Mobile Edge Computing} 
Mobile edge computing (MEC) is regarded as a promising solution to break through the computation limitation~\cite{li2020noma}. Due to the flexibility of the UAVs, the UAV is an ideal mobile edge computing (MEC) platform for performing computing-intensive tasks for ground users. Furthermore, the UAV-enabled MEC platform can be quickly deployed in emergency response scenarios such as major traffic accidents~\cite{zhou2019secure}. There have been several works investigating the performance of UAV-enabled MEC. In~\cite{zhou2018computation}, the authors studied the UAV-enabled MEC wireless powered system by considering both partial and binary computation offloading modes. Instead of using only the UAVs to act as the BSs, the authors in~\cite{8740949} installed the MEC servers on both UAVs and stationary BSs and presented a novel game-theoretic framework to serve their users more efficiently. In~\cite{8933487}, the authors consider both computation bits and energy consumption to optimize the computation efficiency in a multi-UAV MEC system. The authors in~\cite{zhang2019resource} maximize the computation efficiency in partial computation offloading mode.

However, different from the work mentioned above, in this paper, we reduce energy consumption by adopting a CDC technique to mitigate stragglers in the network. The UAVs can recover the computed task if the returned tasks are greater than or equal to the recovery threshold.
\subsection{Stochastic Integer Programming}
Stochastic integer programming is one of the important tools to incorporate uncertainty in optimization problems~\cite{wang2014stochastic}. %\textcolor{blue}{The purpose of stochastic programming is to develop a solution that optimizes some criteria set by the decision maker while also accounting parameters' uncertainty. Because many real-world decisions contain uncertainty,} 
SIP can be applied to various fields to solve the optimization problem, e.g., production planning~\cite{fleming1987optimal}. SIP assumes uncertain data as random variables with known probability distributions, and uses sampled values from this distribution to build a scenario tree and optimize over the expectation~\cite{lara2020electric}. SIP models can correct the decisions using the concept of recourse. In this idea, some decisions have to be made before realizing uncertain parameters and some decisions after their realization~\cite{birge2011introduction}. SIP models can be formulated as the two-stage and multi-stage problems. For the two-stage SIP, stage one decisions are made ‘here and now’ at the beginning of the period without the uncertainty realization. Stage two decisions are taken ‘wait and see’ as the recourse action at the end of the period~\cite{li2020review}. For example, in~\cite{chaisiri2011optimization}, the authors applied the two-stage SIP to optimize the resource provisioning cost in cloud computing. In the courier delivery serves, the authors in~\cite{8108576} uses the two-stage SIP to plan an optimal vehicle delivery route. A multi-stage SIP is a generalization of the two-stage SIP to the sequential realization of uncertainties. For example, the authors in~\cite{liu2017multistage} use a multi-stage SIP to optimize electricity generation, storage, and transmission investments over a long planning horizon. The recourse is the key concept behind SIP. In this problem, weather, demand, and shortfall uncertainties are constantly changing. Therefore, it is not possible to obtain one decision that is suitable for all scenarios. With the idea of recourse, corrective action can be made after a random event has taken place. To the best of our knowledge, the application of stochastic programming to coded distributed computing has been less studied.

\subsection{Coded Distributed Computing}
Distributed computing has been widely adopted to perform various computation tasks in different computing systems~\cite{kartik1997task,lu2019toward}. Nevertheless, there are many design problems, i.e., computing frameworks are vulnerable to uncertain disturbances, such as node failures, communication congestion, and straggler nodes~\cite{wang2019batch}. Only in recent years, coding techniques gained great success in improving the resilience of communication, storage, and cache systems to uncertain system noises~\cite{9024694}. The authors have~\cite{lee2017speeding} first presented the used of CDC to speed up matrix multiplication and data shuffling. As a result, a lot of the focus has been shifted to CDC. Followed by this study, CDC has been explored in many different computation problems, such as the gradients~\cite{tandon2017gradient}, large matrix-matrix multiplication~\cite{lee2017high}, and multivariate polynomials~\cite{yu2019lagrange}.

There have been many other works to reduce the communication load~\cite{li2015coded, 8758338} that are capable of improving the overall communication time. The authors in~\cite{li2015coded} introduced a Coded MapReduce framework to reduce the inter-server communication load by a multiplicative factor that grows linearly with the number of servers in the system. The authors in~\cite{8758338} presented a technique known as Short-Dot to reduces the cost of computation, storage, and communication. Besides reducing the communication load, Short-Dot also tackles the straggler issue. It completes the computation successfully by ignoring the stragglers. More relevant to our study, the authors in~\cite{dutta2019optimal} proposed PolyDot codes, which is a unified view of Matdot~\cite{dutta2019optimal} and Polynomial codes~\cite{yu2017polynomial} and leads to a trade-off between recovery threshold and communication costs.

However, the works mentioned above mainly focus on the designing of different CDC schemes. Therefore, in this paper, we adopt PolyDot codes in the UAV network to alleviate the straggler problem and improve network reliability.

\section{System Model}\label{system_model}
\begin{figure}[t]
    \includegraphics[width=8.5cm, height=10cm,trim={0cm 17cm 9cm 0},clip]{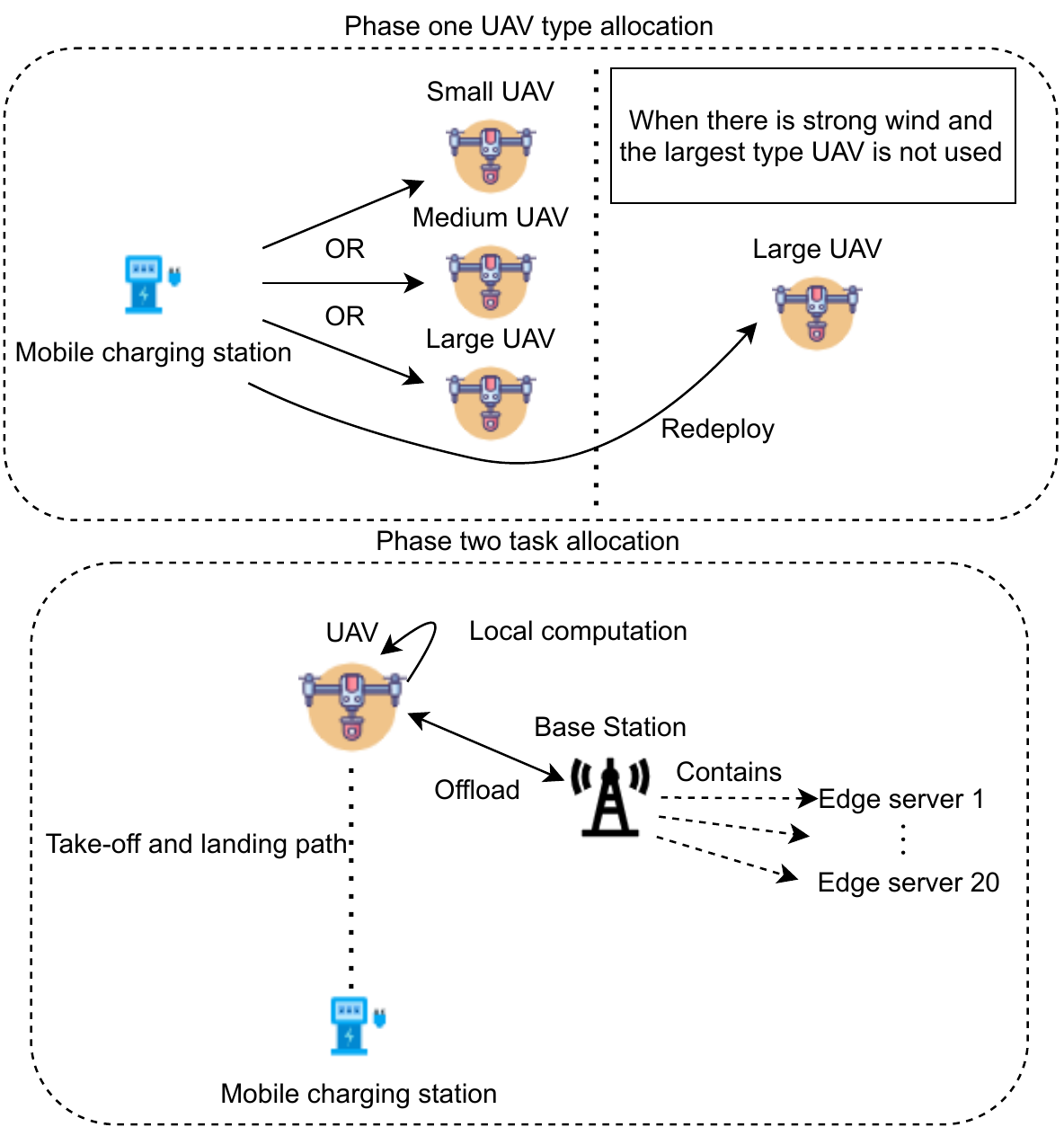}
  \caption{An illustrative example of the network with $\mathcal{X}~=~\{1:small, 2:medium, 3:large\}$, 1 mobile charging station $Y = 1$, 20 edge servers $q_1=20$ attached to 1 BS $F = 1$.}
  \label{fig:UAV network}
\end{figure}

The overall system model is shown in Fig. \ref{fig:UAV network}. We model the phase one (UAV type allocation) and phase two (task allocation) to complete applications defined by an application owner, e.g., road traffic monitoring~\cite{ro2007lessons} while considering various uncertainties. Since each edge server has limited computation capability, by deploy many edge servers at the BS, we can use constraints (53) and (54) from Appendix A to ensure that there will be enough computation resources to support the computation required by each UAV. The following sets are used to denote time slots, UAV types, mobile charging stations, and BSs.
\begin{itemize}
    \item $\mathcal{T}= \{1,\ldots,\bar{t},\ldots,\bar{T}\}$ represents the different time slot.
    \item $\mathcal{\bar{P}}^{\bar{t}}=\{1,\ldots,\bar{p}^{\bar{t}},\ldots,\bar{P}^{\bar{t}}\}$ represents the period in time slot $\bar{t}$.
    \item The available UAVs are clustered into $|\mathcal{X}|$ types denoted by set $\mathcal{X}$, where $\mathcal{X}=\{1,\ldots,x,\ldots,X\}$. Specifically, the type refers to the battery capacity of the UAV in ascending order. For example, $X$ is the largest type UAV that has the most battery and therefore leads to a longer flight time. The UAVs are owned by service provider $\bar{A}_1$. We use $x^{\bar{t}}$ to denote when type $x$ UAV is used in time slot $\bar{t}$.
    \item $\mathcal{Y}~=\{1,\ldots,y,\ldots,Y\}$ represents the UAV mobile charging stations, owned by service provider $\bar{A}_1$. All the mobile charging stations are deployed at pre-specified locations defined by application owner $\bar{A}_3$. 
    \item Each of BS $f$ is attached with $q_f$ number of edge servers. $\mathcal{F}~= \{1,\ldots,f,\ldots,F\}$ represents BSs with the height of $H_f$. Edge servers are owned by service provider $\bar{A}_2$.
\end{itemize}

 In phase one, the application owner first considers the weather uncertainty to pre-allocate the UAV types to each mobile charging station, also known as a UAV depot. Once the phase one optimization is done, all the UAVs will take off from their respective mobile charging stations which are located at $(a_y,b_y)$. $(a_y,b_y)$ are the x-y coordinates of mobile charging station $y$. At time slot $\bar{t}$, type $x$ UAV will take off vertically to the height of $H_{y,x^{\bar{t}}}$ and hover in the sky for purposes such as traffic monitoring. $(a_{y},b_{y},H_{y,x^{\bar{t}}})$ and $(a_f,b_f,H_f)$ are the three-dimensional coordinates of the type $x$ UAV associated with mobile charging station $y$ and edge servers in BS $f$, respectively, where $H_{y,x^{\bar{t}}}~>H_f$ to maintain a line-of-sight (LoS) communication link between type $x$ UAV and edge servers in BS $f$. For simplicity, we assume that UAV maintain a LoS link with the edge servers in the RSUs. Due to the hovering capability, we consider only the rotary-wing UAVs~\cite{zeng2019energy}.

After the UAVs reach their respective heights, they can receive and process computation tasks. In this paper, we consider the task that the type $x$ UAV computes is the matrix-matrix product \textbf{A}\textbf{B} involving the two matrices \textbf{A} and \textbf{B}. However, the UAV has limited computing and storage capability~\cite{hu2018joint}. Therefore, the UAV can choose to offload a portion or the whole matrix multiplication to the edge servers~\cite{hu2018joint}. In phase two, it derives the offloading decision to minimize the overall operation cost by considering the demand and shortfall uncertainties. Note that the key notations used in the paper are listed in Table~\ref{table:symbols}. In the following, we discuss the coded distributed computing model and UAV energy consumption model.

\begin{table*}
\centering
 \caption{List of key notation}
 \begin{tabular}{||c|l||} 
 \hline
 Symbol & Definition \\ [0.5ex] 
 \hline
 $\mathcal{T}$ & Set of time slots while $\bar{t}\in\mathcal{T}$ denotes the time slot index\\
  \hline
 $\mathcal{\bar{P}}^{\bar{t}}$ & Set of periods in $\bar{t}$ while $\bar{p}^{\bar{t}}\in\mathcal{\bar{P}}^{\bar{t}}$ denotes the period index\\
 \hline
 $\mathcal{X}$ & Set of UAV types while $x\in\mathcal{X}$ denotes the UAV type index\\
 \hline
 $\mathcal{Y}$ & Set of mobile charging stations while $y\in\mathcal{Y}$ denotes the mobile charging station index\\
 \hline
 $\mathcal{F}$ & Set of BSs while $f\in\mathcal{F}$ denotes the BS index\\
 \hline
 $k$ & Recovery threshold\\
 \hline
 $z$ & The number of stages in multi-stage SIP\\
 \hline
 $\gamma^{\bar{t}}$ & Set of weather condition scenarios in $\bar{t}$ while $\mu_i^{\bar{t}}\in\gamma^{\bar{t}}$ denotes the weather condition scenario index\\
 \hline
 $\Theta^{\bar{t}}$ & Set of demand scenarios in $\bar{t}$ while $\lambda_i^{\bar{t}}\in\Theta^{\bar{t}}$ denotes the demand scenario index\\
 \hline
 $\Omega^{(\bar{z},\bar{t})}$ & \makecell[l]{Set of shortfall scenarios in stage $\bar{z}-1$, where $2<\bar{z}\leq z$, and $\bar{t}$ while $\omega_i^{(\bar{z},\bar{t})}\in\Omega^{(\bar{z},\bar{t})}$ denotes the shortfall \\scenario index}\\
 \hline
 $T_y^{x^{\bar{t}}}$ & \makecell[l]{Binary variable at time slot $\bar{t}$ for mobile charging station $y$ indicates whether type $x$ UAV is used.}\\
 \hline
 $T_y^{(\bar{t},X)}(\mu_i^{\bar{t}})$ & \makecell[l]{Binary variable at time slot $\bar{t}$ for mobile charging station $y$ indicates whether a correction on-demand type-$X$\\ UAV is used in scenario $\mu_i^{\bar{t}}$, and $X$ represents the largest UAV type.} \\
 \hline
 $M_{f}^{(s)}$ & Binary variable to indicate whether the edge servers in BS $f$ will be used or not\\
 \hline
 $M^{(L,2)}_{y,x^{\bar{t}}}(\lambda_i^{\bar{t}})$ & \makecell[l]{Decision variable represents the number of copies computed locally by the type $x$ UAV that is associated with \\mobile charging station $y$ in stage 2, time slot $\bar{t}$ and scenario $\lambda_i^{\bar{t}}$}\\
 \hline
 $M^{(O,2)}_{y,x^{\bar{t}},f}(\lambda_i^{\bar{t}})$ & \makecell[l]{Decision variable represents the number of copies offloaded to the edge servers in BS $f$ by the type $x$ UAV that is \\associated with mobile charging station $y$ in stage 2, time slot $\bar{t}$ and scenario $\lambda_i^{\bar{t}}$}\\
 \hline
 \vdots & \hspace*{+70mm} \vdots\\
 \hline
  $M_{y,x^{\bar{t}}}^{(L,\hat{z})}(\lambda_i^{\bar{t}},\ldots,\omega_i^{(\bar{z},\bar{t})})$ & \makecell[l]{Decision variable represents the number of copies computed locally by the type $x$ UAV that is associated with \\mobile charging station $y$ in time slot $\bar{t}$, scenario $\lambda_i^{\bar{t}}$, \ldots, scenario $\omega_i^{(\bar{z},\bar{t})}$ and stage $\hat{z}$, where $1<\hat{z}\leq z$}\\
  \hline
  $M_{y,x^{\bar{t}},f}^{(O,\hat{z})}(\lambda_i^{\bar{t}},\ldots,\omega_i^{(\bar{z},\bar{t})})$ & \makecell[l]{Decision variable represents the number of copies offloaded to the edge servers in BS $f$ by the type $x$ UAV that is \\associated with mobile charging station $y$ in time slot $\bar{t}$, scenario $\lambda_i^{\bar{t}}$, \ldots, scenario $\omega_i^{(\bar{z},\bar{t})}$ and stage $\hat{z}$}\\
  \hline
\end{tabular}
\label{table:symbols}
\end{table*}

\subsection{Coded Distributed Computing}
Massive parallelization can speed up matrix multiplication. However, it has a computational bottleneck due to stragglers or faults. Coded computation is introduced to make matrix multiplications resilient to faults and delays, i.e., PolyDot codes~\cite{dutta2019optimal}. In PolyDot codes, the system model typically consists of the followings~\cite{dutta2019optimal}:
\begin{itemize}
    \item Master node receives computation inputs, encodes and distributes them to the worker nodes.
    \item Worker nodes perform pre-determined computations on their respective inputs in parallel.
    \item Fusion node receives outputs from successful worker nodes and decodes them to recover the final output.
\end{itemize}
We consider that the type $x$ UAV is our proposed network's master and fusion node. Each edge server in BS $f$ is the worker and has the computation capability of $\tau_{f,q}$, where $\tau_{f,q}$ denotes the CPU computation capability of the edge server $q$ in BS $f$ (in CPU cycles per second). 

The definitions of copy, successful workers, recovery threshold, shortfall, and demand are given as follows.

\noindent\textbf{Definition 1.} [Copy] a fraction of matrices \textbf{A} and \textbf{B}~\cite{dutta2019optimal}.

\noindent\textbf{Definition 2.} [Successful workers] Workers that finish their computation task and the task is received successfully by the UAV. 

\noindent\textbf{Definition 3.} [Recovery threshold] The recovery threshold is the worst-case minimum number of successful workers required by the UAV to complete the computation~\cite{dutta2019optimal}.

\noindent\textbf{Definition 4.} [Shortfall] There exists a shortfall when the total returned copies from the local computation and from the workers are less than the recovery threshold.

\noindent\textbf{Definition 5.} [Demand] The demand is size of the matrix input $N_y$. It is always different as the input of the matrix multiplication is not always the same.

Following \cite{dutta2019optimal}, two $N_{y} \times N_{y}$ square matrices $\textbf{A}_{y}$ and $\textbf{B}_{y}$ are considered. Note that our model can be applied to other matrices, e.g., non square matrices. Each of matrices $\textbf{A}_{y}$ and $\textbf{B}_{y}$ is sliced both horizontally and vertically. For example, $\textbf{A}_{y}$ is sliced into $\frac{N_{y}}{t} \times \frac{N_{y}}{s}$ matrices and $\textbf{B}_{y}$ is sliced into $\frac{N_{y}}{s} \times \frac{N_{y}}{t}$. We choose $s$ and $t$ such that they satisfy $st~= m$~\cite{dutta2019optimal} and a copy is the $m$-th fractions of matrices \textbf{A} and \textbf{B}. Each edge server has a storage constraint that limits the edge server to store only $m$ fractions of matrices \textbf{A} and \textbf{B}~\cite{dutta2019optimal}. The recovery threshold $k$ is defined~\cite{dutta2019optimal} as:
\begin{equation}\label{1_equ}
    k = t^2(2s-1).
\end{equation}

The processing by the workers may take a longer time when it is currently occupied with some other tasks. Therefore, the processing in the offloaded tasks is perceived to have failed if the duration exceeds the threshold time limit~\cite{dutta2017coded}. To recover the computed task, the sum of returned offloaded copies and locally computed copies must be greater than or equal to recovery threshold $k$.

The decision scenario of phase one and phase two are explained using recovery threshold $k$. In phase one, $|\mathcal{X}|$ scenarios may occur. Mobile charging station $y$ chooses the UAV type to be used. In phase two, three scenarios may occur.
\begin{itemize}\label{scenario}
    \item The UAV can compute all copies locally, $M_{y,x^{\bar{t}}}^{(L)}\geq k$, where $M_{y,x^{\bar{t}}}^{(L)}$ indicates the number of copies that type $x$ UAV from mobile charging station $y$ computes locally at time slot $\bar{t}$ and $k$ is defined in (\ref{1_equ}).
    \item The UAV can offload all copies to BS $f$, $M_{y,x^{\bar{t}},f}^{(O)}\geq k$, where $M_{y,x^{\bar{t}},f}^{(O)}$ denotes the number of copies that are offloaded to the edge servers in BS $f$ at time slot $t$ by the type $x$ UAV from mobile charging station $y$.
    \item The UAV can compute some copies locally and offload some copies to the edge servers in BS $f$, $M_{y,x^{\bar{t}}}^{(L)} + M_{y,x^{\bar{t}},f}^{(O)}\geq k$
\end{itemize}   
The final output can be decoded from all the return copies $M_{y,x^{\bar{t}}}^{(L)}+M_{y,x^{\bar{t}},f}^{(O)}$.

Similar to~\cite{dutta2019optimal}, the type $x$ UAV associated with mobile charging station $y$ uses $d_{enc}=~N_y^2(M_{y,x^{\bar{t}}}^{(L)}+M_{y,x^{\bar{t}},f}^{(O)})$ symbols for encoding of matrices and $d_{dec}~=~N_y^2t^2(2s-1)\log^2t^2(2s-~1)$ to decode the returned matrices. Each copy contains $m$-th fractions of matrices \textbf{A} and \textbf{B}. UAV will transmit $d_{comm}^{to}=~ \frac{N_y^2}{m}$ symbols to each of the edge servers. Each copy requires $d_{cmp} =~ \frac{N_y^3}{mt}$ symbols for computation. After computation is completed, the edge server will send $d_{comm}^{fr}=~\frac{N_y^2}{t^2}$ symbols back to the UAV. 

\subsection{UAV Hovering Energy}
The propulsion energy consumption is needed to provide the UAV with sufficient thrust to support its movement. Note that we drop the time notation for ease of presentation. The propulsion power of a rotary-wing UAV with speed $V$ can be modeled as follows~\cite{zeng2019energy}:
\begin{align}\label{propul_power_equ}
   P_{x}(V) = P_{x,0}\biggl(1+\frac{3V^2}{U^2_{tip}}\biggr)+\hspace*{+35mm}\noindent\nonumber\\P_{x,1}\biggl(\sqrt{1+\frac{V^4}{4v_0^4}}-\frac{V^2}{2v_0^2}\biggr)^\frac{1}{2} + \frac{1}{2}d_0\rho^{\bar{t}} \mathbb{A}V^3,
\end{align}
where
\begin{align}
    &P_{x,0} = \frac{\delta}{8}\rho \mathbf{s}\mathbb{A}\triangle_{x}^3R^3, \\ 
    &P_{x,1} = (1+r)\frac{W_{x}^{3/2}}{\sqrt{2\mathbb{A}\rho}}.
\end{align}
$P_{x,0}$ and $P_{x,1}$ are two constants related to UAV's weight, rotor radius, air density, etc. $U_{tip}$ denotes the tip speed of the rotor blade, $v_0$ is known as the mean rotor induced velocity in hover, $d_0$ and $s$ are the fuselage drag ratio and rotor solidity, respectively. $\rho$ and $\mathbb{A}$ are the air density and rotor disc area, respectively. $r$ is the incremental correction factor to induced power. $W_x$ is the type $x$ UAV weight, $\delta$ is the profile drag coefficient, and $\triangle_x$ denotes blade angular velocity of the type $x$ UAV. By substituting $V=0$ into~(\ref{propul_power_equ})~\cite{zeng2019energy}, we obtain the power consumption for hovering status as follows:
\begin{equation}
    P_{x,h} = P_{x,0}+P_{x,1}.
\end{equation}

\subsection{Local Computing Model}
When one copy is processed locally, the local computation execution time of the type $x$ UAV is expressed as~\cite{pham2021uav}:
\begin{equation}
    t_{y,x}^{local}(N_y) = \frac{\mathcal{C}_xd(\frac{N_y^3}{mt})}{\tau_x},
\end{equation}
where $\mathcal{C}_x$ is the number CPU cycles needed to process a bit, $\tau_x$ denotes the total CPU computing capability of the type $x$ UAV, and $d(\cdot)$ is a function to translate the number of symbols to the number of bits for computation, i.e., if the 16 Quadrature Amplitude Modulation (QAM) is used, each symbol carries 4 bits \cite{qam}. The type $x$ UAV takes $t_{y,x}^{enc}$ seconds to encode one copy of the matrices, and it is expressed as follows:
\begin{equation}
    t_{y,x}^{enc}(N_y) = \frac{\mathcal{C}_xd(N_y^2)}{\tau_x}.
\end{equation}

After the type $x$ UAV obtains at least $k$ copies, it will take $t_{y,x}^{dec}$ seconds to decode. $t_{y,x}^{dec}$ is defined as follows:
\begin{equation}
    t_{y,x}^{dec}(N_y) = \frac{\mathcal{C}_xd(N_y^2t^2(2s-1)\log^2t^2(2s-1))}{\tau_x}.
\end{equation}

\subsection{UAV Communication Model}
We assume that each UAV is allocated with an orthogonal spectrum resource block to avoid the co-interference among the UAVs~\cite{zhou2019energy}. The transmission rate from the type $x$ UAV which is associated with mobile charging station $y$ to the edge servers in BS $f$ can be represented as~\cite{chen2020intelligent}:
\begin{equation}\label{transmission_rate}
    r_{y,x,f} = B_{x}\log_2(1+P_{x}^Ch_{y,x,f}/N_o),
\end{equation}
where the wireless transmission power of the type $x$ UAV at time slot $\bar{t}$ is expressed as $P_{x}^C$ and $B_{x}$ is the bandwidth. $h_{y,x,f}$ is the channel gains, and $N_0$ is the variance of complex white Gaussian noise. The UAV to edge server communication is most likely to be dominated by LoS links. Therefore, the air-to-ground channel power gain from the type $x$ UAV to the edge servers in BS $f$ can be modeled as follows~\cite{hua2019energy}:
\begin{equation}\label{gain}
    h_{y,x,f} = \frac{\beta_0}{\mathbb{D}_{y,x,f}^2},
\end{equation}
where
\begin{equation}
    \mathbb{D}_{y,x,f}^2 = (a_y-a_f)^2+(b_y-b_f)^2 + (H_{y,x}-H_f)^2.
\end{equation}
$\mathbb{D}_{y,x,f}$ denotes the distance between the type $x$ UAV that is associated with mobile charging station $y$ and the edge servers in BS $f$, and $\beta_0$ represents the reference channel gain at distance $d_0 = 1$m in an urban area~\cite{hua2019energy}. We assume that for all the edge servers in the same BS $f$, they will have the same $\mathbb{D}_{y,x,f}^2$. The transmission time to offload one copy of matrix from the type $x$ UAV to a edge server in BS $f$ can be given as follows:
\begin{equation}\label{transmission_time}
    t_{y,x,f}^{to}(N_y) = \frac{d(\frac{N_y^2}{m})}{r_{y,x,f}}.
\end{equation}
The energy $e_{y,x}$ required by the type $x$ UAV to receive data from the edge server in BS $f$ is defined as follows~\cite{ng2020joint}:
\begin{align}
    e_{y,x}(N_y) = P_{x}^{re}\frac{d(\frac{N_y^2}{t^2})}{r_{f,y,x}},
\end{align}
where $P_{x}^{re}$ is the receiving power of type $x$ UAV. $r_{f,y,x}$ is the transmission rate from edge servers in BS $f$ to type $x$ UAV which is associated with mobile charging station $y$. It is define similar to~(\ref{transmission_rate}).
%Average hover time is 14 minutes \cite{muntaha2020energy}.

\begin{figure*}[htp]
    \centering
    \includegraphics[width=16cm, height=3cm,trim={0cm 26cm 0cm 0},clip]{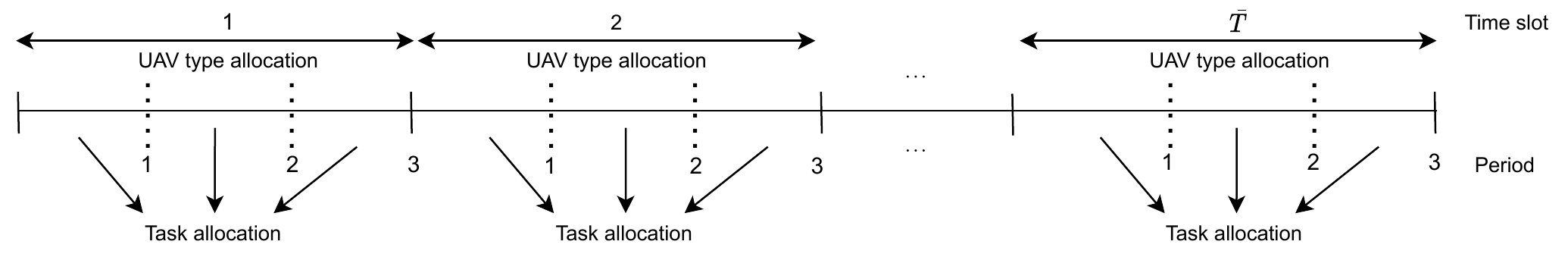}
  \caption{The decision process of the system across all the time slots $\bar{T}$.}
  \label{fig:time}
\end{figure*}
\begin{figure*}[htp]
    \centering
    \includegraphics[width=14cm, height=12cm,trim={0cm 18cm 6cm 0},clip]{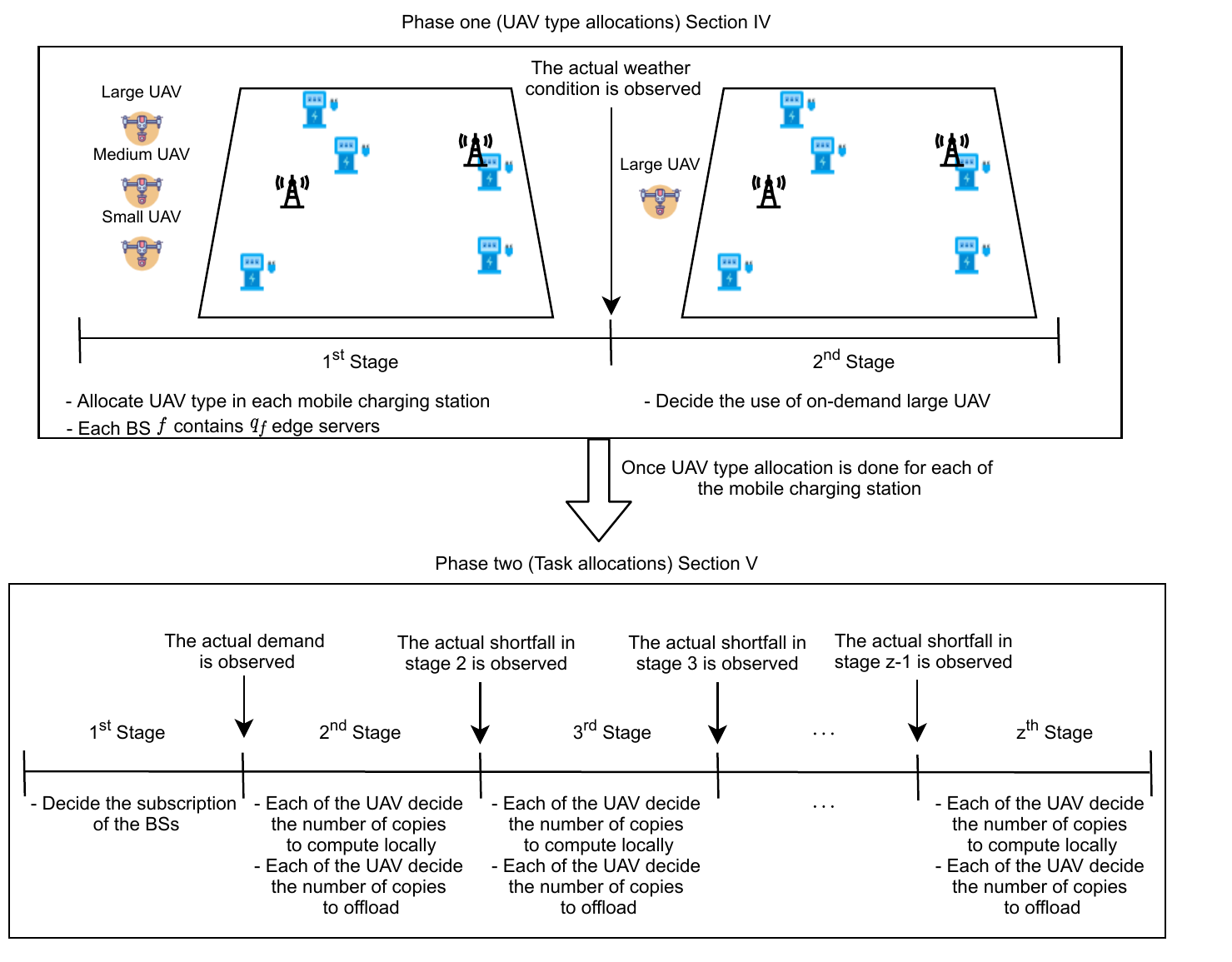}
  \caption{Decision making process of the system in one time slot with the using of three different types of UAV, $\mathcal{X}~=~\{1:small, 2:medium, 3:large\}$.}
  \label{fig:decision}
\end{figure*}
\subsection{Problem Formulation}
As an illustration, Fig.~\ref{fig:time} depicts the decision process of the system across all the time slots. UAV type allocation is performed in each time slot $\bar{t}$. Throughout $\bar{t}$, the mobile charging stations will use the same UAV type to perform the task allocation in each period. Fig.~\ref{fig:decision} shows a detailed diagram that zooms into one-period in one-time slot, and it is explained in details in both Sections~\ref{UAV allocation} and~\ref{Task allocation}. In Section~\ref{UAV allocation}, the application owner $\bar{A}_3$ pays a reservation cost to make an advance booking for a different time slot for the use of the UAVs. The application owner can observe the weather condition via weather forecast as it may affect the status of the UAV. If the wind is too strong and the UAV used is not large type, the UAV may crash as it has insufficient energy to hover against the wind~\cite{bezzo2016online}. For example, a strong wind has high kinetic energy, kinetic energy leads to a higher density of the air, and it increases the UAV hovering power consumption. Low wind speed is referred to as wind speed that is less than $11m/s$ and turbulence level $<5$~\cite{chu2021simulation}. As a result, $\bar{A}_3$ has to request an on-demand $X$ type UAV to perform the job. In order for SCOS to model the weather uncertainty, we formulate the two-stage SIP to optimize the UAV type allocation.

To achieve cost minimization, phase two (task allocation) in Section \ref{Task allocation} has to consider two sources of uncertainty, i.e., the demand uncertainty and shortfall uncertainty. Demand uncertainty refers to the task required by the applications, such as traffic monitoring can be of different sizes, i.e., the task's size depends on the image resolution. Shortfall uncertainty refers to if the UAV offloads the copies to the edge servers, the computed copies may not return, or the number of copies returned is less than the recovery threshold due to delays and link failure. Therefore, we use multi-stage SIP to model the demand uncertainty to optimize the number of copies to compute locally $M_{y,x^{\bar{t}}}^{(L)}$ and offload $M_{y,x^{\bar{t}},f}^{(O)}$.  For example, when the recovery threshold is $k=4$ and the UAV decide to offload two copies of the task for the edge servers in BS $f$ to compute, i.e., $M_{y,x^{\bar{t}},f}^{(O)}=~2$. Therefore, the UAV has to compute at least two more copies locally to match the recovery threshold $M_{y,x^{\bar{t}}}^{(L)}=~2$. In time slot $\bar{t}$, type $x$ UAV will hover in the sky for a threshold time limit $t^{thresh}_{y,x^{\bar{t}}}$ to wait for the offloaded copies to return. Without loss of generality, $t^{thresh}_{y,x^{\bar{t}}}$ is set as the worst-case scenario, i.e., the time required to compute all copies locally by the UAV, i.e.,~$M_{y,x^{\bar{t}}}^{L}=k$. However, there is a probability that the edge servers in BSs may fail, i.e., the computed task is not returned to the UAV before $t^{thresh}_{y,x^{\bar{t}}}$. As a result, the UAV cannot complete the full task if the total returned copies are less than 4. When the UAV fails to receive sufficient number of copies, there are shortfalls, and hence, the UAV has to re-compute the shortfalls locally or re-offload to the edge servers. Since the UAV has limited computation capabilities, it can choose to re-compute the shortfall locally or re-offload to the edge servers until $z$ stages, where $z$ is the number of times of re-computations. In the meantime, the UAV has to continue hover in the sky when performing the re-computation. In order to model the shortfall uncertainty, we formulate $z$-stage SIP to optimize the numbers of copies to compute locally and to offload, and we can also optimize the number of stages required. Hence, this scheme will minimize the overall network cost, and the system model of this network is formulated as follows:

\noindent $\displaystyle\min_{T_{y}^{x^{\bar{t}}},T_{y}^{x^{\bar{t}}}(\mu_i^{\bar{t}}), M_{f}^{(s)},\ldots,M_{y,x^{\bar{t}},f}^{(O,\hat{z})}(\lambda_i^{\bar{t}},\ldots,\omega_i^{(\bar{z},\bar{t})})}$:
\begin{align}
    \sum_{\bar{t}\in\mathcal{\bar{T}}}\mathcal{O}^{UAV}_{allocation}(\bar{t}) + \sum_{\bar{t}\in\mathcal{\bar{T}}}\sum_{\bar{p}^{\bar{t}}\in\mathcal{\bar{P}}^{\bar{t}}}\mathcal{O}^{Task}_{allocation}(\bar{p}^{\bar{t}}),
\end{align}
subject to: (\ref{2cons1})-(\ref{2cons4}), (\ref{cons1})-(\ref{cons13})

\noindent where $\mathcal{O}^{UAV}_{allocation}(\bar{t})$ is the UAV type allocation cost in time slot $\bar{t}$ and it is defined in (\ref{2sip_1}) in Section \ref{UAV allocation}. $\mathcal{O}^{Task}_{allocation}(\bar{p}^{\bar{t}})$ is the task allocation cost within period $\bar{p}$ and period $\bar{p}$ is in time slot $\bar{t}$. The task allocation is defined in (\ref{sip_1}) in Section \ref{problem}.

\section{Phase one: UAV type allocation}\label{UAV allocation}
This section introduces the SIP to optimize phase one (UAV type allocation) in SCOS by minimizing the total allocation. As described in Section~\ref{system_model}, the application owner $\bar{A}_3$ needs to make a reservation in advance to secure certain types of UAVs, which are own by $\bar{A}_1$. However, the weather condition is unknown and may vary at a different time slot $\bar{t}$. If the wind is too strong, the UAV is required to use more energy to hover at a fixed location~\cite{bezzo2016online}. As a result, the UAV will crash with insufficient energy, and the application owner has to make an on-demand request with a $X$ type UAV. Fig.~\ref{fig:decision} illustrates the decision-making process of the system with the use of three UAV types, 1, 2 and 3, which represents small, medium and large, respectively.

Hence, we formulate this scheme as the two-stage SIP model.
\begin{itemize}
    \item \textbf{First stage:} The application owner makes a reservation on the types of UAVs to be used. The decision will be made based on the available cost information and the probability distribution of the weather condition.
    \item \textbf{Second stage:} After knowing the exact weather condition, the application owner decides the correction action, which is the on-demand request to use the largest type $X$ UAV.
\end{itemize}

Let $\mu_i^{\bar{t}} = \{G_1^{\bar{t}}(\mu_i^{\bar{t})},\ldots,G_y^{\bar{t}}(\mu_i^{\bar{t}})$\} denote weather condition scenarios $i$ of all mobile charging stations at time slot $\bar{t}$. The set of all weather scenarios is denoted by $\gamma^{\bar{t}}$, i.e., $\mu^{\bar{t}}_i\in~\gamma^{\bar{t}}$~\cite{8482480}. $G^{\bar{t}}_y(\mu_i^{\bar{t}})$ represents a binary parameter of the weather condition at time slot $\bar{t}$. For tractability, we only consider that each mobile charging station experiences only two types of weather condition. As shown in Table~\ref{table:weather}, $G^{\bar{t}}_y(\mu_i^{\bar{t}})=1$ means that at time slot $\bar{t}$, the wind is strong in mobile charging station $y$ and the UAV has crashed, and $G^{\bar{t}}_y(\mu_i^{\bar{t}})=0$ means otherwise. $\mathcal{P}(\mu^{\bar{t}})$ denotes the probability if scenario $\mu^{\bar{t}}\in\gamma^{\bar{t}}$ is realized. All of the scenarios can be obtained from historical records~\cite{8108576} or weather forecast.
\begin{table}
\centering
 \caption{Weather uncertainty}
 \begin{tabular}{||c|l||} 
 \hline
 Symbol & Definition \\ [0.5ex] 
 \hline
 $G^{\bar{t}}_y(\mu_i^{\bar{t}})=0$ & \makecell[l]{At time slot $\bar{t}$, the wind is weak or there is no wind \\at mobile charging station $y$. Low wind speed is\\ referring to wind speed that is less than 11 $m/s$\\ and turbulence level
$<5$~\cite{chu2021simulation}}\\
  \hline
   $G^{\bar{t}}_y(\mu_i^{\bar{t}})=1$ & \makecell[l]{At time slot $\bar{t}$, the wind is strong at mobile charging \\station $y$. High wind speed is referring to wind \\speed that is greater than 11 $m/s$ and \\turbulence level
 $>5$~\cite{chu2021simulation}}\\
  \hline
\end{tabular}
\label{table:weather}
\end{table}
The cost function is proportional to the resources used. In total, there are $|\mathcal{X}|+2$ types of payments. Note that we drop the time notation.
\begin{itemize}
    \item $C_r^x$ is the reservation cost for the type $x$ UAV. It is defined as follows:
    \begin{equation}
     C_r^x = \alpha_1 B_x,
    \end{equation}
    where $B_x$ is the battery capacity of the type $x$ UAV and $\alpha_1$ is the cost coefficient.
    \item $C_o^{X}$ is the on-demand cost for the type $X$ UAV, which represents the largest UAV type. It is defined as follows:
    \begin{equation}
     C_o^{X} = \alpha_2 B_{X},
    \end{equation}
    where $\alpha_2$ is the cost coefficient with a similar role to $\alpha_1$ and $\alpha_2>\alpha_1$. $B_{X}$ is the battery capacity of the type $X$ UAV.
    \item $C_p$ is the penalty cost. This penalty cost is the repair cost for the crashed UAV.
\end{itemize}

We formulate the UAV type allocation as a two-stage SIP model. There are $|\mathcal{T}|(|\mathcal{X}|+1)$ decision variables in this model. 
\begin{itemize}
    \item $T_y^{x^{\bar{t}}}$ is a binary variable at time slot $\bar{t}$ for mobile charging station $y$ indicates whether type $x$ UAV is used. When $T_y^{x^{\bar{t}}} = 1$, at time slot $\bar{t}$, mobile charging station $y$ uses type $x$ UAV and $T_y^{x^{\bar{t}}}=0$ means otherwise.
    
    \item $T_y^{(\bar{t},X)}(\mu_i^{\bar{t}})$ is a binary variable at time slot $\bar{t}$ for mobile charging station $y$ indicates whether a correction on-demand type $X$ UAV is used in scenario $\mu_i^{\bar{t}}$, and $X$ represents the largest UAV type. When $T_y^{(\bar{t},X)}(\mu_i^{\bar{t}}) = 1$, at time slot $\bar{t}$, mobile charging station $y$ performs a correction action by using the largest type-$X$ UAV in scenario $\mu_i^{\bar{t}}$ and $T_y^{(\bar{t},X)}(\mu_i^{\bar{t}})=0$ means otherwise.

\end{itemize}

The objective function given in (\ref{2sip_1}) and (\ref{2sip_2}) is to minimize the cost of the UAV type allocation. The expressions in (\ref{2sip_1}) and (\ref{2sip_2}) represent the first- and second-stage SIP, respectively. The SIP formulation can be expressed as follows:

\noindent $\displaystyle\min_{T_{y}^{x^{\bar{t}}},T_y^{(\bar{t},X)}(\mu_i^{\bar{t}})}$:
\begin{align}\label{2sip_1}
    \sum_{\bar{t}\in\mathcal{T}}\mathcal{O}^{UAV}_{allocation}(\bar{t}) =\sum_{\bar{t}\in\mathcal{T}}\sum_{y\in\mathcal{Y}}\sum_{x\in\mathcal{X}}T_{y}^{x^{\bar{t}}}C_r^x
    +\mathbb{E}\biggl[\mathcal{Q}(T_{y}^{x^{\bar{t}}}(\mu_i^{\bar{t}}))\biggr],
\end{align}
where
\begin{align}\label{2sip_2}
    \mathcal{Q}(T_{y}^{x^{\bar{t}}}(\mu_i^{\bar{t}})) = \sum_{\bar{t}\in\mathcal{T}}\sum_{\mu_i^{\bar{t}}\in\gamma^{\bar{t}}}\mathcal{P}(\mu_i^{\bar{t}})\sum_{y\in\mathcal{Y}}T_y^{(\bar{t},X)}(\mu_i^{\bar{t}})(C_o^{X}+C_p),
\end{align}
subject to: 
\begin{align}
    \sum_{x\in\mathcal{X}}T_{y}^{x^{\bar{t}}} = 1, \hspace*{+28mm}\forall\bar{t}\in\mathcal{T},\forall y\in\mathcal{Y},\hspace*{-0mm}\label{2cons1}
\end{align}
\begin{align}
    \sum_{x\in\mathcal{X}\setminus \{X\}}T_{y}^{x^{\bar{t}}}(1-G_y(\mu_i^{\bar{t}})) + T_{y}^{X^{\bar{t}}} + T_y^{(\bar{t},X)}(\mu_i^{\bar{t}}) = 1, \noindent\nonumber\\ \hspace*{+10mm}\forall\bar{t}\in\mathcal{T},\forall\mu_i^{\bar{t}}\in\gamma^{\bar{t}},\forall y\in\mathcal{Y},\label{2cons2}
\end{align}
\begin{align}
    &T_y^{x^{\bar{t}}}\in\{0,1\}, & \forall\bar{t}\in\mathcal{T},\forall y\in \mathcal{Y}, \forall x\in\mathcal{X},\label{2cons3}\\
    &T_y^{(\bar{t},X)}(\mu_i^{\bar{t}}) \in\{0,1\}, & \forall\bar{t}\in\mathcal{T},\forall y\in \mathcal{Y},\forall\mu_i^{\bar{t}}\in\gamma^{\bar{t}}.\label{2cons4}
\end{align}

\noindent The constraint in (\ref{2cons1}) ensures that the application owner makes a reservation on the types of UAV. On the other hand, (\ref{2cons2}) ensures that the UAV crashes because of strong wind if the application owner previously reserves a UAV that is not largest type $X$. Then, the application owner has to perform a correction action by using the largest type $X$ on-demand UAV. (\ref{2cons3}) and (\ref{2cons4}) are boundary constraints for the decision variables.

\section{Phase two: task allocation}\label{Task allocation}
 Once the types of the UAVs are optimized from phase one in SCOS, we introduce the Deterministic Integer Programming (DIP) and SIP to optimize phase two (the number of copies to compute locally and to offload) by minimizing the UAV network cost. Note that for simplicity, we drop notation $\bar{p}^{\bar{t}}$ from phase two task allocation.

\begin{figure}[t]
    \centering
    \includegraphics[width=8.5cm, height=5cm,trim={1cm 22cm 4cm 0},clip]{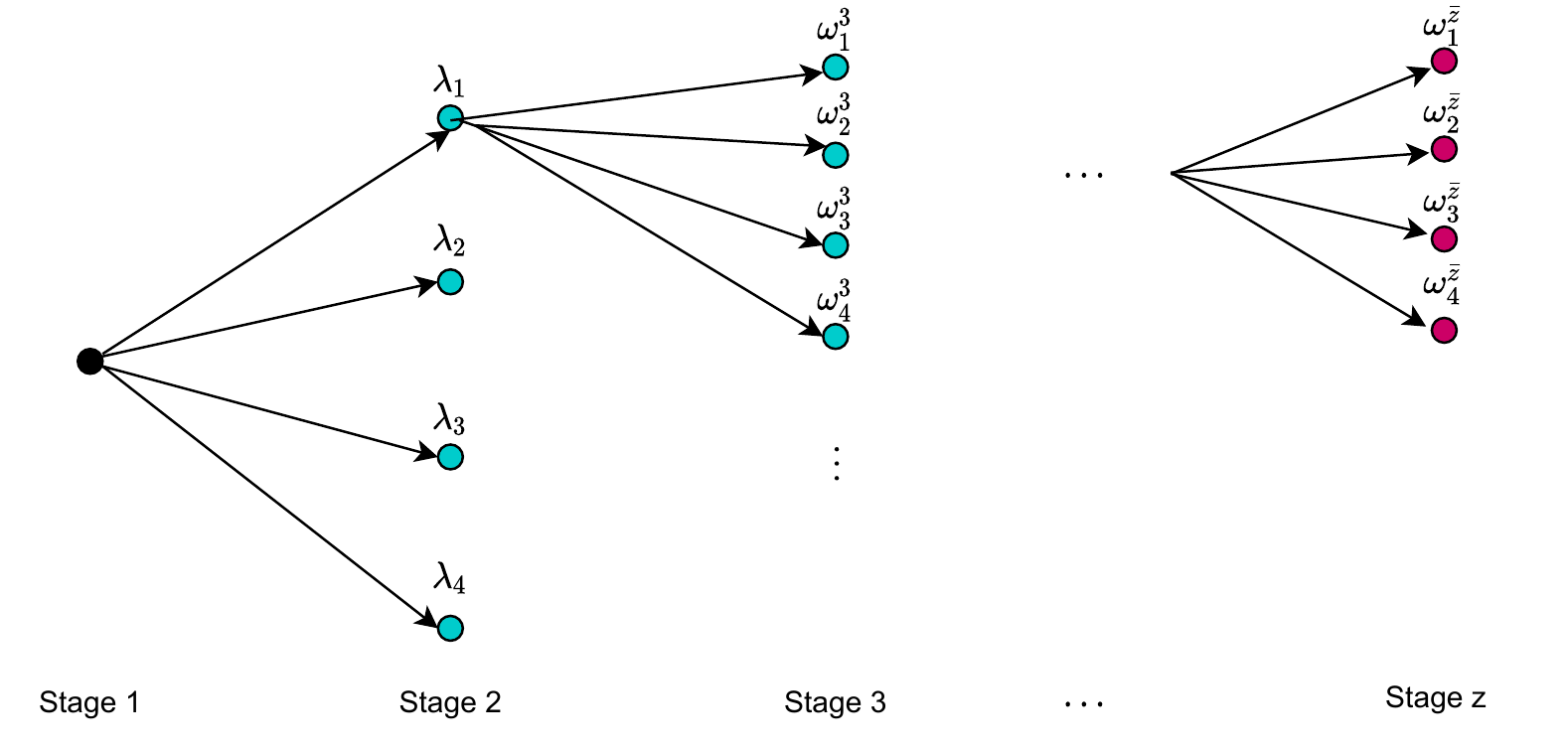}
    \caption{A scenario tree structure for $z$-stage SIP in task allocation.}
    \label{fig:tree}
\end{figure}

\subsection{Deterministic Integer Programming System Model}\label{DIP}
In an ideal case, when the actual demand, which is the actual matrix size and the number of shortfalls, are precisely known ex-ante, the UAVs can choose the exact number of copies to compute locally or offload. Therefore, the correction for the shortfall is not needed, and the correction cost is zero. Similar to~\cite{mitsis2020data}, the cost function is proportional to the UAVs offloaded data and to their demand for consuming computation resources. Choosing different sizes of UAVs will affect the payment value. In total, five types of payments are considered in DIP.
\begin{itemize}
    \item $\Bar{C}_{f}$ is the subscription cost for the edge servers in BS $f$.
    \item $\bar{C}_{y,x}$ denotes the UAV local computation cost and encoding cost for computing of one copy, i.e., 
    \begin{equation}
        \bar{C}_{y,x}(\mathcal{D}) = \alpha_3(t_{y,x}^{local}(\mathcal{D})+t_{y,x}^{enc}(\mathcal{D})),
    \end{equation}
    where $\alpha_3$ is the cost coefficient associated to the energy consumption and $\mathcal{D}$ is the actual demand. 
    \item $C_{y,x^{\bar{t}},f}$ denotes the offloading cost and it consists of three parts. The first part is related to the transmission $t_{y,x^{\bar{t}},f}^{to}$ and encoding delay $t_{y,x}^{enc}$. The second part is the type $x$ UAV energy consumption cost and the last part $C_{f}$ is the service cost for edge servers in BS $f$. It is modeled as follows:
    \begin{align}\label{transmission_cost}
        C_{y,x^{\bar{t}},f}(\mathcal{D}) = \alpha_3(t_{y,x^{\bar{t}},f}^{to}(\mathcal{D})+t_{y,x}^{enc}(\mathcal{D})) \nonumber\\+\alpha_4e_{y,x^{\bar{t}}}(\mathcal{D}) + C_{f},
    \end{align}
    where $\alpha_4$ is the cost coefficient with a similar role to $\alpha_3$.
    \item $\underbar{C}^{thresh}_{y,x^{\bar{t}}}$ denotes the hovering cost for $t^{thresh}_{y,x^{\bar{t}}}$ seconds. They are defined as follows:
    \begin{align}
        &\underbar{C}^{thresh}_{y,x^{\bar{t}}} = t_{y,x^{\bar{t}}}^{thresh}k\alpha_5P_{x^{\bar{t}},h},
    \end{align}
    where $\alpha_5$ is the cost coefficient with a similar role to $\alpha_3$.
    \item $\hat{C}_{y,x}$ denotes the type $x$ UAV decoding cost for the returned matrices as follows:
    \begin{equation}
        \hat{C}_{y,x}(\mathcal{D}) = \alpha_3t_{y,x}^{dec}(\mathcal{D}).
    \end{equation}
\end{itemize}

A DIP can be formulated and minimize the total cost of the UAVs as follows:

\noindent $\displaystyle\min_{M_{f}^{(s)},\ldots,M_{y,x^{\bar{t}},f}^{(O)}}$:
\begin{align}\label{dip}
    \sum_{\bar{t}\in\mathcal{T}}\mathcal{O}^{Task}_{allocation}(\bar{t}) =\sum_{\bar{t}\in\mathcal{T}}\sum_{f\in\mathcal{F}}M_{f}^{(s)}\Bar{C}_{f}+\nonumber\noindent\\ \sum_{\bar{t}\in\mathcal{T}}\sum_{y\in \mathcal{Y}}\sum_{f\in\mathcal{F}}\biggl(M_{y,x^{\bar{t}}}^{(L)}\bar{C}_{y,x}(\mathcal{D}) +  M^{(TH)}_{y,x^{\bar{t}},f}\underbar{C}_{y,x}^{thresh}+ \nonumber\noindent\\
    M_{y,x^{\bar{t}},f}^{(O)}C_{y,x,f}(\mathcal{D}) + \hat{C}_{y,x}(\mathcal{D}) \biggr),
\end{align}
subject to:
\begin{align}
    &\sum_{y\in\mathcal{Y}}M_{y,x^{\bar{t}},f}^{(O)} \leq \sigma M_f^{(s)}, &\forall\bar{t}\in\mathcal{T},\forall f \in\mathcal{F},\label{dip_cons_1}\\
    &\sum_{y\in\mathcal{Y}}M_{y,x^{\bar{t}},f}^{(O)} \leq q_{f}, &\forall\bar{t}\in\mathcal{T},\forall f \in\mathcal{F},\label{dip_cons_2}\\
    &M_{y,x^{\bar{t}},f}^{(O)} \leq \sigma M^{(TH)}_{y,x^{\bar{t}},f}, &\forall y\in \mathcal{Y},\forall\bar{t}\in\mathcal{T},\forall f \in\mathcal{F},\label{dip_cons_3}\\
    &\sum_{f\in\mathcal{F}}M_{y,x^{\bar{t}},f}^{(O)} \geq S_{y}^{\bar{t}} - M_{y,x^{\bar{t}}}^{(L)}, &\forall\bar{t}\in\mathcal{T},\forall y\in \mathcal{Y},\label{dip_cons_4}
\end{align}
\begin{align}
    M_{y,x^{\bar{t}}}^{(L)} + \sum_{f\in\mathcal{F} }M_{y,x^{\bar{t}},f}^{(O)} -(S_{y}^{\bar{t}}-M_{y,x^{\bar{t}}}^{(L)})\geq k,\hspace*{+10mm}\nonumber\\\forall y\in \mathcal{Y}, \forall\bar{t}\in\mathcal{T},\forall f\in \mathcal{F},\label{dip_cons_5}
\end{align}
\begin{align}
    &M_f^{(s)},M^{(TH)}_{y,x^{\bar{t}},f} \in\{0,1\}, & \forall y\in \mathcal{Y}, \forall f \in\mathcal{F},\label{dip_cons_6}\\
    &M_{y,x^{\bar{t}}}^{(L)}, M_{y,x^{\bar{t}},f}^{(O)} \in \{0,1,\ldots\}, &\forall\bar{t}\in\mathcal{T},\forall y\in \mathcal{Y}, \forall f\in \mathcal{F}.\label{dip_cons_7}
\end{align}

\noindent $M_{f}^{(s)}$ is a binary variable to indicate whether the edge servers in BS $f$ will be used or not. $M^{(TH)}_{y,x^{\bar{t}},f}$ is a binary variable to indicate in time slot $\bar{t}$ whether the type $x$ UAV which is associated with mobile charging station $y$ will choose to offload or not. When $M^{(TH)}_{y,x^{\bar{t}},f}=1$, in time slot $\bar{t}$ the UAV associated with mobile charging station $y$ choose to offload some of the copies to BS $f$ and $M^{(TH)}_{y,x^{\bar{t}},f}=0$ means otherwise. The objective function in~(\ref{dip}) is to minimize UAVs' total cost involving the UAVs' local computation cost and the UAVs' offloading cost. The constraint in~(\ref{dip_cons_1}) ensures that the subscription cost of the edge servers in the BS will be paid if they are used in any of the stages, where $\sigma$ is a sufficiently large number. (\ref{dip_cons_2}) ensures that the total number of copies offloaded to the edge servers must not exceed the total number of edge servers in BSs. (\ref{dip_cons_3}) ensures that the threshold cost will be paid if the UAV perform offloading action. (\ref{dip_cons_4}) ensures that the shortfalls should only exist if the number of copies offloaded is more than or equal to the shortfalls. (\ref{dip_cons_5}) ensures that the number of copies computed locally and offloaded have to be at least equal to or larger than recovery threshold $k$. (\ref{dip_cons_6}) indicates $M_f^{(s)}$ and $M^{(TH)}_{y,x^{\bar{t}},f}$ are binary variables. (\ref{dip_cons_7}) indicates that $M_{y,x^{\bar{t}}}^{(L)}$ and $M_{y,x^{\bar{t}},f}^{(O)}$ are positive decision variables.

\subsection{Stochastic Integer Programming System Model}\label{problem}
This section introduces the SIP to minimize the total cost of the network by optimizing the number of copies to compute locally and to offload to the edge servers in BSs. The first stage consists of all decisions that have to be selected before the demand and shortfall are realized and observed. In the second stage and onwards, decisions are allowed to adapt to this information. In each stage, decisions are limited by constraints that may depend on previous decisions and observations.

As described in Section~\ref{system_model}, there is a subscription cost when the service provider $\bar{A}_3$ wants to use the edge servers in BSs for computation. Then, without knowing the demand, the type $x$ UAV can decides the number of copies to compute locally $M_{y,x^{\bar{t}}}^{(L)}$ and the number of copies to offload $M_{y,x^{\bar{t}},f}^{(O)}$.

The computation process in the edge servers are not very reliable, as the edge servers might be processing some other task or congested. As a result, the computation time is much longer than the threshold limit $t^{thresh}_{y,x^{\bar{t}}}$. Therefore, if a copy is offloaded, there is a probability that the computation might fail, and it will require the type $x$ UAV to re-offload again or compute it locally.

Hence, we formulate this framework as a $z$-stage SIP model.
\begin{itemize}
    \item \textbf{First stage:} The application owner $\bar{A}_3$ decides to use the edge servers in BS $f$ or not. The decision will be made based on the available cost information, the probability distribution of the demand, and the shortfall.
    \item \textbf{Second stage:} After knowing the exact demand, the application owner $\bar{A}_3$ decides the number of copies that are computed locally and the number of copies to be offloaded to the edge servers in BS $f$.
    \item \textbf{Third stage:} After knowing the exact shortfall in the previous stage, the $\bar{A}_3$ performs a correction action to re-decide the number of copies that is computed locally and the number of copies to be offloaded to the edge servers in BS $f$.
    \\\hspace*{+40mm}\vdots
    \item \textbf{$z$ stage:} After knowing the exact shortfall in the $z$-1 stage, $\bar{A}_3$ performs a correction action to re-decide the number of copies that is computed locally and the number of copies to be offloaded to the edge servers in BS $f$. To promote the UAV to complete the task, a huge penalty will occur if there is still a shortfall in stage $z$.
\end{itemize}

Let $\lambda_i^{\bar{t}} = \{D_1^{\bar{t}},\ldots,D_y^{\bar{t}}\}$ denote the UAV demand scenario $i$ across all mobile charging station $y$ in time slot $\bar{t}$ and the set of demand scenarios is denoted by $\Theta^{\bar{t}}$, i.e., $\lambda_i^{\bar{t}}\in\Theta^{\bar{t}}$~\cite{8482480}. $D_y^{\bar{t}}$ contains a discrete value from a finite set $\mathcal{W} = \{1,\ldots,W\}$, it represents the size of the task in UAV that is associated with mobile charging station $y$. Specifically, $D_y^{\bar{t}}=1000$ means that in time slot $\bar{t}$ the matrix that UAV receives is in the size of $1000\times1000$. Let $\omega^{(\bar{z},\bar{t})}_i =\{\mathbb{F}_{1}^{(\bar{z},\bar{t})},\ldots,\mathbb{F}_{y}^{(\bar{z},\bar{t})}\}$ denote the $i$-th shortfall scenario of the UAV in time slot $\bar{t}$ that is associated with its individual mobile charging station in stage $\bar{z}-1$, where $2<\bar{z}\leq z$. The set of shortfall scenarios is denoted by $\Omega^{(\bar{z},\bar{t})}$, i.e., $\omega^{(\bar{z},\bar{t})}_i\in \Omega^{(\bar{z},\bar{t})}$. $\mathbb{F}_{y}^{(\bar{z},\bar{t})}$ represents a binary parameter of the shortfall in time slot $\bar{t}$ from the type $x$ associated with mobile charging station $y$ in stage $\bar{z}-1$. For example, $\mathbb{F}_y^{(\bar{z},\bar{t})} = 1$ means that, in time slot $\bar{t}$ from the copies that the UAV has offloaded, at least $A_y^{(\bar{z},\bar{t})}$ copy did not return. As a result, the total number of copies that the UAV currently has is less than $k$, and $\mathbb{F}_y^{(\bar{z},\bar{t})} = 0$ means otherwise. In stage $z$, $\mathbb{F}_y^{(\bar{z},\bar{t})} = 0$ when $\mathbb{F}_y^{(\bar{z}-1,\bar{t})} = 0$. When there is no shortfall in the previous stage then, there will not be any shortfall in the next stage. Fig.~\ref{fig:tree} illustrates the stages with four scenarios at each stage. All of the scenarios can be obtained from the historical records.

%From the commercial perspective, the service provider of the RSUs want to attracting more customers. Therefore it carry out some marketing strategies, one of which is that the services can be sold in a reservation way with a cheaper price and a real-time/on-demand requested way with an expensive price.

The cost function used in SIP is similar to DIP with an additional penalty cost $\Tilde{C}$. $\Tilde{C}$ occurs when the UAV still has to perform a corrective action. In total, six types of payments are considered in $z$-stage SIP. 

We formulate the task allocation as the $z$-stage SIP model. There are $|\bar{\mathcal{T}}|(z(f+2)-1)$ decision variables in this model. 
\begin{itemize}
    \item $M_{f}^{(s)}$ is a binary variable to indicate whether the edge servers in BS $f$ will be used or not. When $M_{f}^{(s)} = 1$, edge servers in BS $f$ will be used and $M_{f}^{(s)}=0$ means otherwise.
    \item $M_{y,x^{\bar{t}},f}^{(O,2)}(\lambda_i^{\bar{t}})$ indicates in time slot $\bar{t}$ the number of copies to be offloaded to the edge servers in BS $f$ by type $x$ UAV which is associated with mobile charging station $y$ in stage 2.
    \item $M_{y,x^{\bar{t}}}^{(L,2)}(\lambda_i^{\bar{t}})$ indicates in time slot $\bar{t}$ the number of copies computed locally by the type $x$ UAV which is associated with mobile charging station $y$ in stage 2.
    \item $M^{(TH,3)}_{y,x^{\bar{t}},f}(\lambda_i^{\bar{t}})$ is a binary variable to indicate in time slot $\bar{t}$ whether the type $x$ UAV which is associated with mobile charging station $y$ choose to offload or not in stage 3. When $M^{(TH,3)}_{y,x^{\bar{t}},f}(\lambda_i^{\bar{t}})=1$, the type $x$ UAV associated with mobile charging station $y$ chooses to offload some of the copies to the edge servers in BS $f$ and $M^{(TH,3)}_{y,x^{\bar{t}},f}(\lambda_i^{\bar{t}})=0$ means otherwise.
    \\\hspace*{+40mm}
    \vdots
    \item $M_{y,x^{\bar{t}},f}^{(O,\hat{z})}(\lambda_i^{\bar{t}},\ldots,\omega_i^{(\bar{z},\bar{t})})$ indicates in time slot $\bar{t}$ the number of copies to be offloaded by type $x$ UAV which is associated with mobile charging station $y$ to the edge servers in BS $f$ in stage $\hat{z}$, where $1<\hat{z}\leq z$.
    \item $M_{y,x^{\bar{t}}}^{(L,\hat{z})}(\lambda_i^{\bar{t}},\ldots,\omega_i^{(\bar{z},\bar{t})})$ indicates in time slot $\bar{t}$ the number of copies that the type $x$ UAV which is associated with mobile charging station $y$ is computing locally in stage $\hat{z}$.
    \item $M^{(TH,\hat{z})}_{y,x^{\bar{t}},f}(\lambda_i^{\bar{t}},\ldots,\omega_i^{(\bar{z},\bar{t})})$ indicates is a binary variable to indicate in time slot $\bar{t}$ whether the type $x$ UAV which is associated with mobile charging station $y$ choose to offload or not in stage $\hat{z}$.
\end{itemize}

The objective function given in~(\ref{sip_1})~-~(\ref{sip_4}) is to minimize the total cost of the network. The expressions in~(\ref{sip_1}), ~(\ref{sip_2}), ~(\ref{sip_3}) and ~(\ref{sip_4}) represent the first, second, third and up till $z$ stage objectives, respectively. $\mathcal{P}(\lambda_i^{\bar{t}})$ and $\mathcal{P}(\omega_i^{(\bar{z},\bar{t})})$ denote the probabilities if scenarios $\lambda_i^{\bar{t}}\in\Theta^{\bar{t}}$ and $\omega_i^{(\bar{z},\bar{t})}\in\Omega^{(\bar{z},\bar{t})}$ are realized, respectively. The SIP formulation can be expressed as follows:

\noindent $\displaystyle\min_{M_{f}^{(s)},\ldots,M_{y,x^{\bar{t}},f}^{(O,\hat{z})}(\lambda_i^{\bar{t}},\ldots,\omega_i^{(\bar{z},\bar{t})})}$:
\begin{align}\label{sip_1}
    \sum_{\bar{t}\in\mathcal{T}}\mathcal{O}^{Task}_{allocation}(\bar{t}) = \sum_{\bar{t}\in\mathcal{T}}\sum_{f\in\mathcal{F}}M_{f}^{(s)}\Bar{C}_{f}
    +\mathbb{E}\biggl[\mathcal{Q}(M_f^{(s)}(\lambda^{\bar{t}}_i))\biggr],
\end{align}
where
\begin{align}\label{sip_2}
    \mathcal{Q}(M_f^{(s)}(\lambda^{\bar{t}}_i)) = \sum_{\bar{t}\in\mathcal{T}}\sum_{\lambda_i^{\bar{t}}\in\Theta^{\bar{t}}}\mathcal{P}(\lambda_i^{\bar{t}})\sum_{y\in \mathcal{Y}}\sum_{f\in\mathcal{F}}\biggl(M_{y,x^{\bar{t}}}^{(L,2)}(\lambda_i^{\bar{t}})\nonumber\noindent\\\bar{C}_{y,x}(D_y(\lambda_i^{\bar{t}}))   +
    M_{y,x^{\bar{t}},f}^{(O,2)}(\lambda_i^{\bar{t}})C_{y,x,f}(D_y(\lambda_i^{\bar{t}})) + \nonumber\noindent\\\underbar{C}_{y,x}^{thresh} + \hat{C}_{y,x}(D_y(\lambda_i^{\bar{t}})) +\nonumber\noindent\\
    \mathbb{E}\biggl[\mathcal{Q}(M_{y,x^{\bar{t}},f}^{(O,2)}(\lambda_i^{\bar{t}},\omega_i^{(3,\bar{t})}))\biggr]\biggr),
\end{align}
\begin{align}\label{sip_3}
    \mathcal{Q}(M_{y,x^{\bar{t}},f}^{(O,2)}(\lambda_i^{\bar{t}},\omega_i^{(3,\bar{t})})) = 
    \sum_{\bar{t}\in\mathcal{T}}\sum_{\omega_i^{(3,\bar{t})}\in\Omega^{(3,\bar{t})}}\mathcal{P}(\omega_i^{(3,\bar{t})})
    \sum_{y\in\mathcal{Y}}\sum_{f\in\mathcal{F}}\nonumber\noindent\\\biggl( 
    M_{y,x^{\bar{t}}}^{(L,3)}(\lambda_i^{\bar{t}},\omega_i^{(3,\bar{t})})
    \bar{C}_{y,x}(D_y(\lambda_i^{\bar{t}})) + M_{y,x^{\bar{t}},f}^{(O,3)}(\lambda_i^{\bar{t}},\omega_i^{(3,\bar{t})})\nonumber\noindent\\C_{y,x,f}(D_y(\lambda_i^{\bar{t}})  + M^{(TH,3)}_{y,x^{\bar{t}},f}(\lambda_i^{\bar{t}},\omega_i^{(3,\bar{t})})\underbar{C}_{y,x}^{thresh}  \nonumber\noindent\\ + \mathbb{E}\biggl[\mathcal{Q}(M_{y,x^{\bar{t}},f}^{(O,3)})(\lambda_i^{\bar{t}},\omega_i^{(3,\bar{t})},\omega_i^{(4,\bar{t})})\biggr]\biggr),\\
    \vdots\hspace*{44mm}\nonumber\noindent
\end{align}
\begin{align}\label{sip_4}
    \hspace*{0mm}\mathcal{Q}(M_{y,x^{\bar{t}},f}^{(O,\hat{z})})(\lambda_i^{\bar{t}},\omega_i^{(3,\bar{t})},\ldots,\omega_i^{(\bar{z},\bar{t})}) =
    \sum_{\bar{t}\in\mathcal{T}}\sum_{\omega_i^{(\bar{z},\bar{t})}\in\Omega^{(\bar{z},\bar{t})}}\nonumber\noindent\\\mathcal{P}(\omega_i^{(\bar{z},\bar{t})})
    \sum_{y\in\mathcal{Y}}\sum_{f\in\mathcal{F}}
    \biggl(M_{y,x^{\bar{t}}}^{(L,\hat{z})}(\lambda_i^{\bar{t}},\ldots,\omega_i^{(\bar{z},\bar{t})})\bar{C}_{y,x}(D_y(\lambda_i^{\bar{t}})) + \nonumber\noindent\\ M^{(TH,\bar{z})}_{y,x^{\bar{t}},f}(\lambda_i^{\bar{t}})\underbar{C}_{y,x}^{thresh} + M_{y,x^{\bar{t}},f}^{(O,\hat{z})}(\lambda_i^{\bar{t}},\ldots,\omega_i^{(\bar{z},\bar{t})})
    \nonumber\noindent\\C_{y,x,f}(D_y(\lambda_i^{\bar{t}})) +\mathbb{F}^{\bar{z}}_y(\lambda_i^{\bar{t}},\ldots,\omega_i^{(\bar{z},\bar{t})})\Tilde{C}\biggr),
\end{align}
subject to: Please see Appendix~\ref{append}

\noindent The constraints in~(\ref{cons1}) and~(\ref{cons2}) are the same as the constraint~(\ref{dip_cons_1}). (\ref{cons20}) and~(\ref{cons21}) are the same as~(\ref{dip_cons_3}). (\ref{cons5}) and~(\ref{cons6}) ensure that if the previous stage does not have a shortfall, and then the shortfall for the next stage should be zero. The shortfalls should only exist if the number of copies offloaded in the previous stage is more than or equal to the shortfall. They also ensure that the number of shortfalls can be reduced when the UAV performs local computations. (\ref{cons7}) ensures that the number of copies computed locally and offloaded should be at least $k$. (\ref{cons8}) and~(\ref{cons9}) ensure that if the UAV has a shortfall in the previous stage, the UAV has to compute the shortfall locally or re-offload to the edge servers in the BS to match $k$, i.e., the recovery threshold for task completion. (\ref{cons10}) and~(\ref{cons11}) ensure that the total number of copies offloaded to the edge servers in each stage must not exceed the total number of edge servers in BSs. (\ref{cons12}) indicates $M_f^{(s)}$,  $M^{(TH,3)}_{y,x^{\bar{t}},f}(\lambda_i^{\bar{t}},\omega_i^{(3,\bar{t})})$, $\ldots,M^{(TH,\bar{z})}_{y,x^{\bar{t}},f}(\lambda_i^{\bar{t}},\ldots,\omega_i^{(\bar{z},\bar{t})})$ are binary variables. (\ref{cons13}) indicates $M_{y,x^{\bar{t}},f}^{(O,2)}(\lambda_i^{\bar{t}})$, $M_{y,x^{\bar{t}}}^{(L,2)}(\lambda_i^{\bar{t}}),\ldots,$  $M_{y,x^{\bar{t}},f}^{(O,\hat{z})}(\lambda_i^{\bar{t}},\ldots,\omega_i^{(\bar{z},\bar{t})}),$ $M_{y,x^{\bar{t}}}^{(L,\hat{z})}(\lambda_i^{\bar{t}},\ldots,\omega_i^{(\bar{z},\bar{t})})$ are positive decision variables. 

To solve SIP, we assume that the probability distribution of all scenarios in set $\gamma^{\bar{t}}$, $\Theta^{\bar{t}}$, $\Omega^{(3,\bar{t})},\ldots,\Omega^{(\bar{z},\bar{t})}$ are known~\cite{dyer2006computational}, then, the complexity of the problem increases exponentially when the total number of scenarios across all the stages increases~\cite{dyer2006computational,rajendran2017platelet}. For phase one in Section~\ref{UAV allocation}, the total number of decision variables can be calculated from (\ref{one_dec}), where $t'$, $y'$, and $x'$ denote the total number of time slots, mobile charging stations, and UAV types in the set, respectively. $\mu'$ is the total number of weather scenarios. The total number of constraints can be calculated from (\ref{one_con}). For example, with six time slots, six mobile charging stations, three UAV types, and ten weather scenarios, phase one will have 468 decision variables and 864 constraints.

\begin{equation}\label{one_dec}
    decisionVariables1 = t'y'x' + \mu't'y'
\end{equation}
\begin{equation}\label{one_con}
    constraints1 = t'y' + 2\mu't'y' + t'y'x'
\end{equation}

For phase two in Section~\ref{problem}, the total number of decision variables can be calculated from (\ref{two_dec}), where $f'$ denotes the total number of BSs in the set. $\lambda'$, $\omega^{3'}$, \ldots, $\omega^{\bar{z}'}$ are the total number of demand, shortfall in stage 2, \ldots, shortfall in stage $\bar{z}-1$ scenarios. The total number of constraints can be calculated from (\ref{two_con}).
\begin{align}\label{two_dec}
    decisionVariables2 = t'f' + t'\lambda'y'f' + t'\omega^{3'}y'f' + \ldots \\\nonumber+ t'\omega^{\bar{z}'}y'f'
\end{align}
\begin{align}\label{two_con}
    constraints2 = 2(t'f'\lambda'\omega^{3'} + \ldots + t'f'\lambda'\omega^{3'}\ldots\omega^{\bar{z}'} \\\nonumber  t'f'\lambda') 
    +2(t'y'f'\lambda'\omega^{3'} + \ldots + t'y'f'\lambda'\omega^{3'}\ldots\omega^{\bar{z}'})\\\nonumber 
    +t'y'\lambda'\omega^{3'} + \ldots + t'y'\lambda'\omega^{3'}\ldots\omega^{\bar{z}'}\\\nonumber 
    +t'y'f'\lambda' + 2(t'y'f'\lambda'\omega^{3'}\ldots\omega^{\bar{z}'})
\end{align}
\subsection{Stochastic Coded Offloading Scheme Flowchart}
\begin{figure}[htp]
    \centering
    \includegraphics[width=8cm, height=7cm,trim={0cm 20cm 10cm 0},clip]{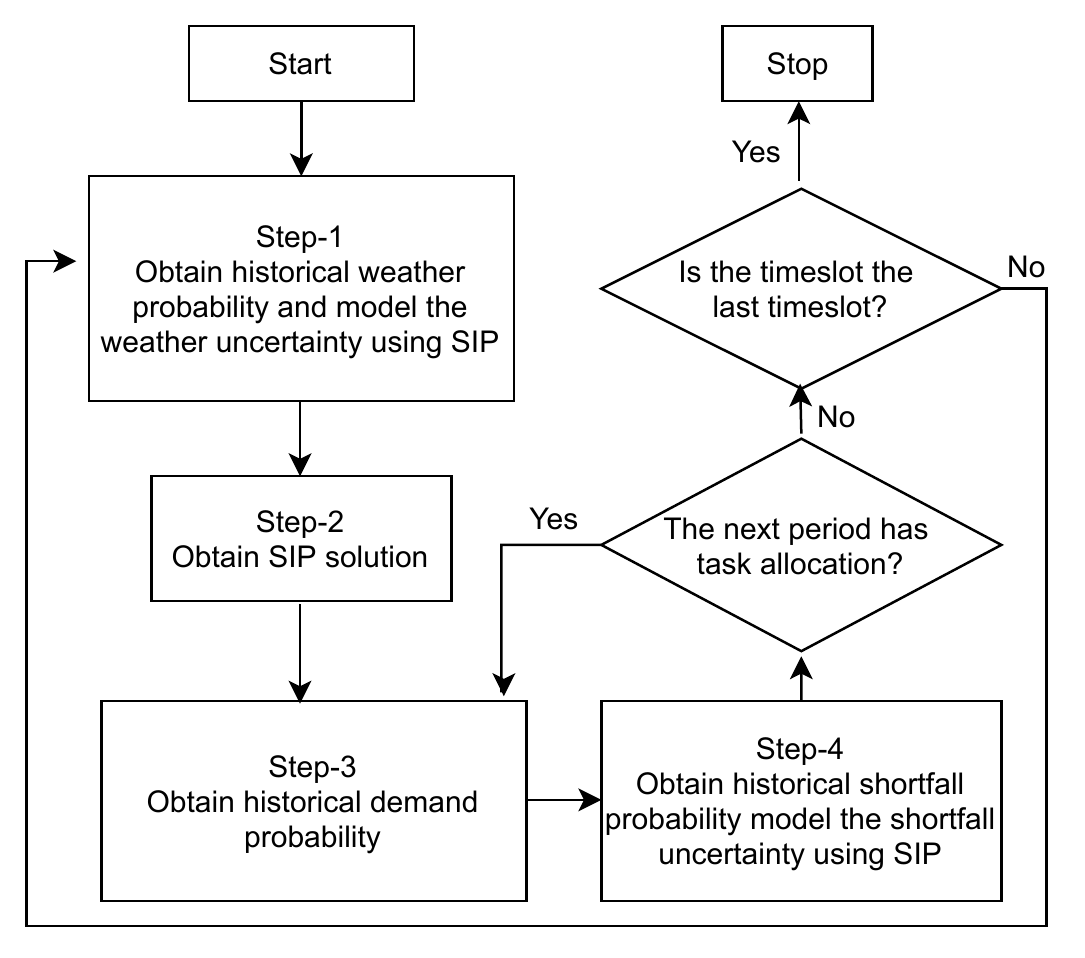}
  \caption{Flowchart of SCOS algorithm.}
  \label{fig:flowchart}
\end{figure}

In Fig.~\ref{fig:flowchart}, the flowchart of SCOS algorithm is shown. The algorithm for solving SCOS is presented in four steps (i.e., Step-1 to Step-4) as follows.

In Step-1, we obtain the weather probability for the first time slot, and it can be obtained from the historical records~\cite{8108576} or weather forecast. The weather uncertainty is then modeled using SIP from~(\ref{2sip_1}).

In Step-2, we save the solution from Step-1.

In Step-3, we obtain the demand probability from the road traffic data set released by Land Transport Authority Singapore~\cite{data}. 

In Step-4, we obtain the shortfall probability from the historical records. Using the allocated UAV type from Step-2, we modeled the demand probability from Step-3 and the shortfall probability using SIP from~(\ref{sip_1}). After solving this problem, if the next period has task allocation, the solution is saved, and the algorithm will proceed to Step-3. Otherwise, the algorithm proceeds to the next decision box. If the timeslot is the last time slot for the next decision box, the algorithm will end and output the SIP decision variables solution. Otherwise, the algorithm will proceed to Step-1.

\begin{figure*}[ht]
\centering
\begin{multicols}{2}
\includegraphics[width=7cm, height=5cm,trim={1cm 18cm 5.5cm 0cm},clip]{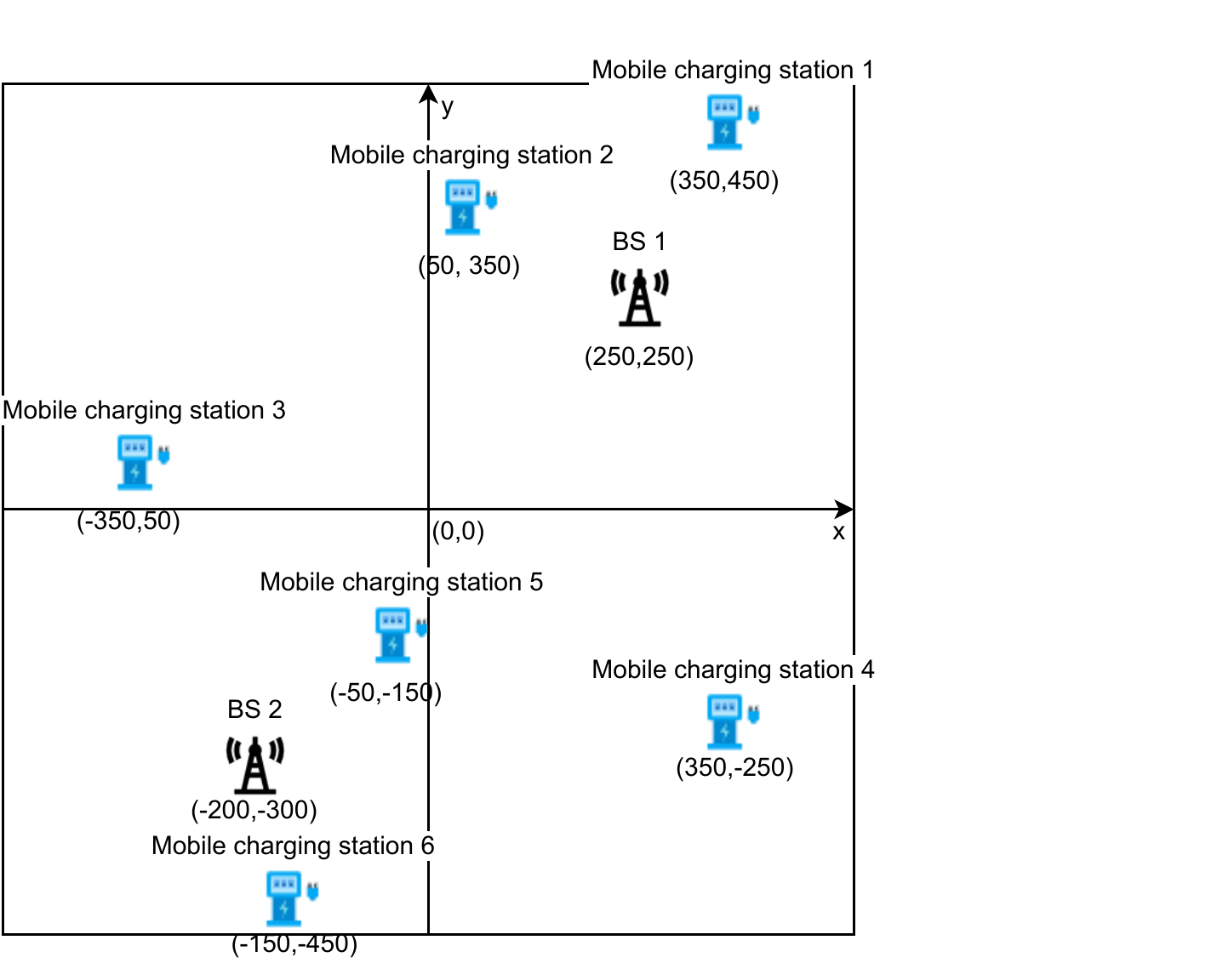}
\caption{x-y coordinates of all the UAVs and BSs.}
\label{fig:location}
\includegraphics[width=8.5cm, height=5cm,trim={5.5cm 18.3cm 5.5cm 5cm},clip]{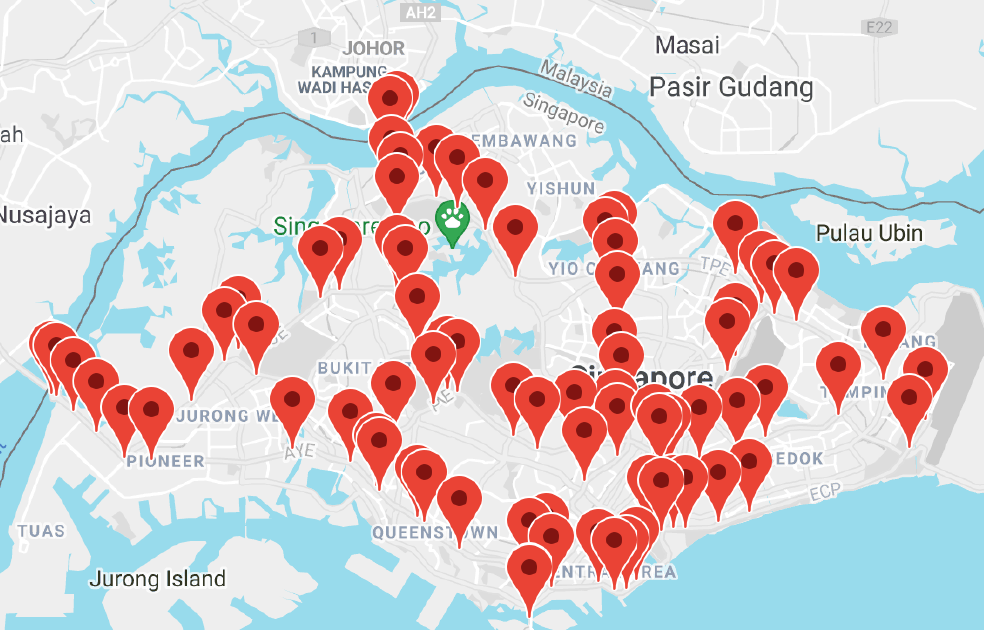}\par
\caption{Singapore traffic camera location.}
\label{fig:traffic camera}
\end{multicols}
\end{figure*}
\begin{figure}[t]
\centering
\includegraphics[width=0.7\columnwidth]{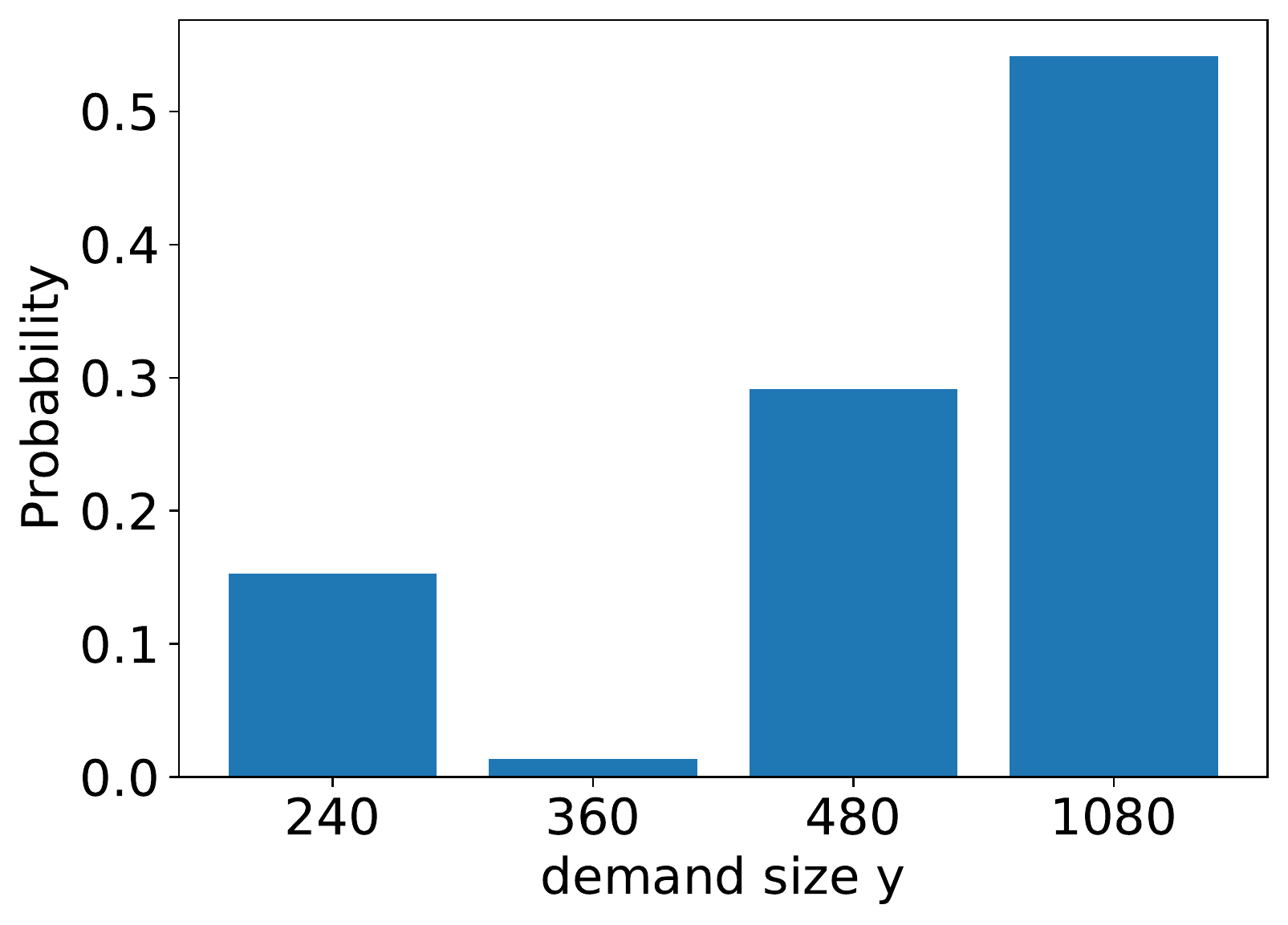}
\caption{The probability distribution of the demand size.}
\label{fig:probability dis}
\end{figure}
\section{Simulation result and analysis}\label{simulation}
In this simulation, we consider the system model with one time slot $|\mathcal{T}|=1$, six mobile charging stations, two BSs, sixty edge servers and thirty UAVs equally split into three UAV types. $\mathcal{X} = \{1:small, 2:medium, 3:large\}$. The battery capacities $B_{1}$, $B_{2}$ and $B_{3}$ are $2375mAh$, $3500mAh$ and $5200mAh$, respectively~\cite{UAV}. A conceptual illustration of the system model is shown in Fig.~\ref{fig:location}. Fig.~\ref{fig:location} is an x-y plane that shows the locations of the UAVs and the BSs. All the UAVs associated with their respective mobile charging stations are hovering at 100m, and the height of all edge servers in the BSs is 20m. They are randomly allocated in the area of $1000\times1000$~$m^2$. Each small grid is $25\times25$~$m^2$. We consider the case with $m = 2$~\cite{dutta2019optimal}. Therefore, we can substitute $s=\frac{2}{t}$ into~(\ref{1_equ}), and we obtain the following:
\begin{equation}\label{diff}
    k = 4t-t^2,
\end{equation}
by differentiating~(\ref{diff}) with respect to the variable $t$ and equating the result to zero. Then, we are able to obtain $t=2$, $s=1$, and $k = 4$. The simulation parameters are summarized in Table~\ref{table:table1} and their values are from~\cite{chen2020intelligent, zeng2019energy}. For the presented experiments, we implement the SIP model using GAMS script~\cite{chattopadhyay1999application,GamsSoftware2013}.

\begin{table}
\centering
\caption{Experiment parameters}
\begin{tabular}{|l| c|} 
 \hline
  \textbf{Parameter} & \textbf{Values} \\ 
 \hline\hline
 UAV weight in $kg$, $W_x$                                 & [8-12]  \\ 
 \hline
 Air density in $kg/m^3$, $\rho$                         & 1.225\\
 \hline
 Rotor radius in meter, $R$                              & 0.5\\
 \hline
 Rotor disc area in $m^2$, $\mathbb{A}$                  & 0.79\\
 \hline
 Blade angular velocity in radians/second, $\triangle_x$   & [380-420]\\
 \hline
 Tip speed of the rotor blade, $U_{tip}$                 & 200\\
 \hline
 Number of blades, $b$                                   & 4\\
 \hline
 Blade or aerofoil chord length, $c$                     & 0.0196\\
 \hline
 \makecell[l]{Rotor solidity, defined as the ratio of the \\ 
 total blade area $bcR$ to the disc area $\mathbb{A}$, $\mathbf{s}$}   & 0.05\\
 \hline
 Fuselage drag ratio, $d_0$                              & 0.3\\
 \hline
 Mean rotor induced velocity in hover, $v_0$             & 7.2\\
 \hline
 Profile drag coefficient, $\delta$                      & 0.012\\
 \hline
 \makecell[l]{Incremental correction factor \\
 to induced power, $r$}
            & 0.1\\
 \hline
 %\makecell[l]{UAV waiting time for the return\\ of offloaded copies in second, %$t^{thresh}_y$}
 %           & 5\\
 UAV hover height in meter, $H_{y,x}$                        & [80-100]\\
 \hline
 Height of edge servers in BS, $H_f$                               & 20\\
 \hline
 UAV bandwidth in $MHz$, $B_x$                           & 2\\
 \hline
 UAV transmit power in $mW$, $P_x^C$                     & 32\\
 \hline
 UAV receiving power in $mW$, $P_x^{re}$                    & 32\\
 \hline
 White  Gaussian  noise in $dBm$, $N_0$                  & -100 \\
 \hline
 UAV computation power in $GHz$, $\tau_x$                & [0.6-1]\\
 \hline
 \makecell[l]{Number of CPU cycles needed\\ to process a bit, $\mathcal{C}_x$}
 & 20\\
 \hline
 Channel gains, $\beta_0$                               & -60dB\\
\hline
 $C_{f}$ in \$                                        & 0.05\\
 \hline
 $\alpha_1$                                              & 0.001\\
 \hline
  $\alpha_2$                                             & 0.0015\\
  \hline
  $\alpha_3$                                             & 0.5\\
  \hline
  $\alpha_4$                                             & 0.5\\
  \hline
 $\alpha_5$                                             & 0.0001\\
\hline
\end{tabular}
\label{table:table1}
\end{table}

\begin{figure*}[ht]
\centering
\begin{multicols}{3}
\includegraphics[width=5.7cm, height=4cm,trim={2cm 24cm 11.5cm 0cm},clip]{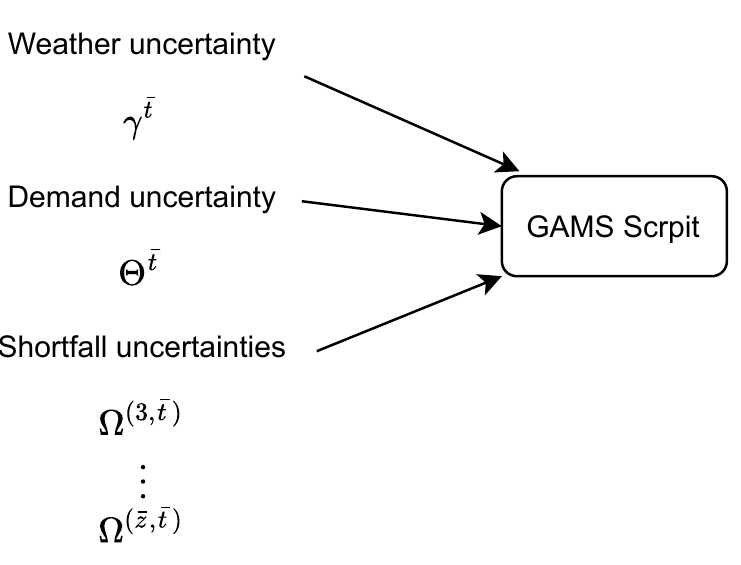}
\caption{The value of uncertainties are inserted into GAMS script.}
\label{fig:insert}
\includegraphics[width=0.8\columnwidth]{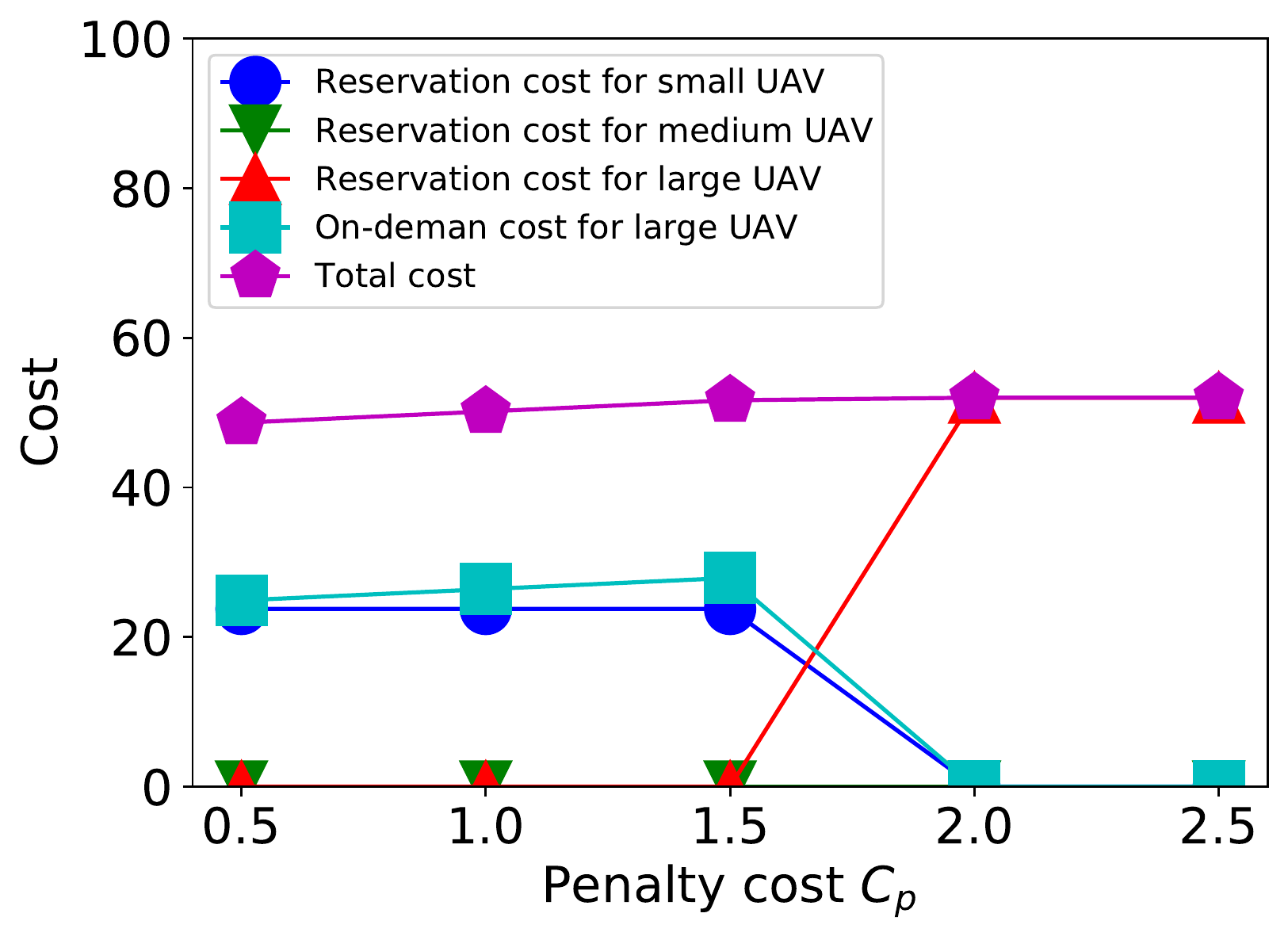}\par
\caption{The cost of UAV type allocation when varying the penalty cost.}
\label{fig:penalty}
\includegraphics[width=0.8\columnwidth]{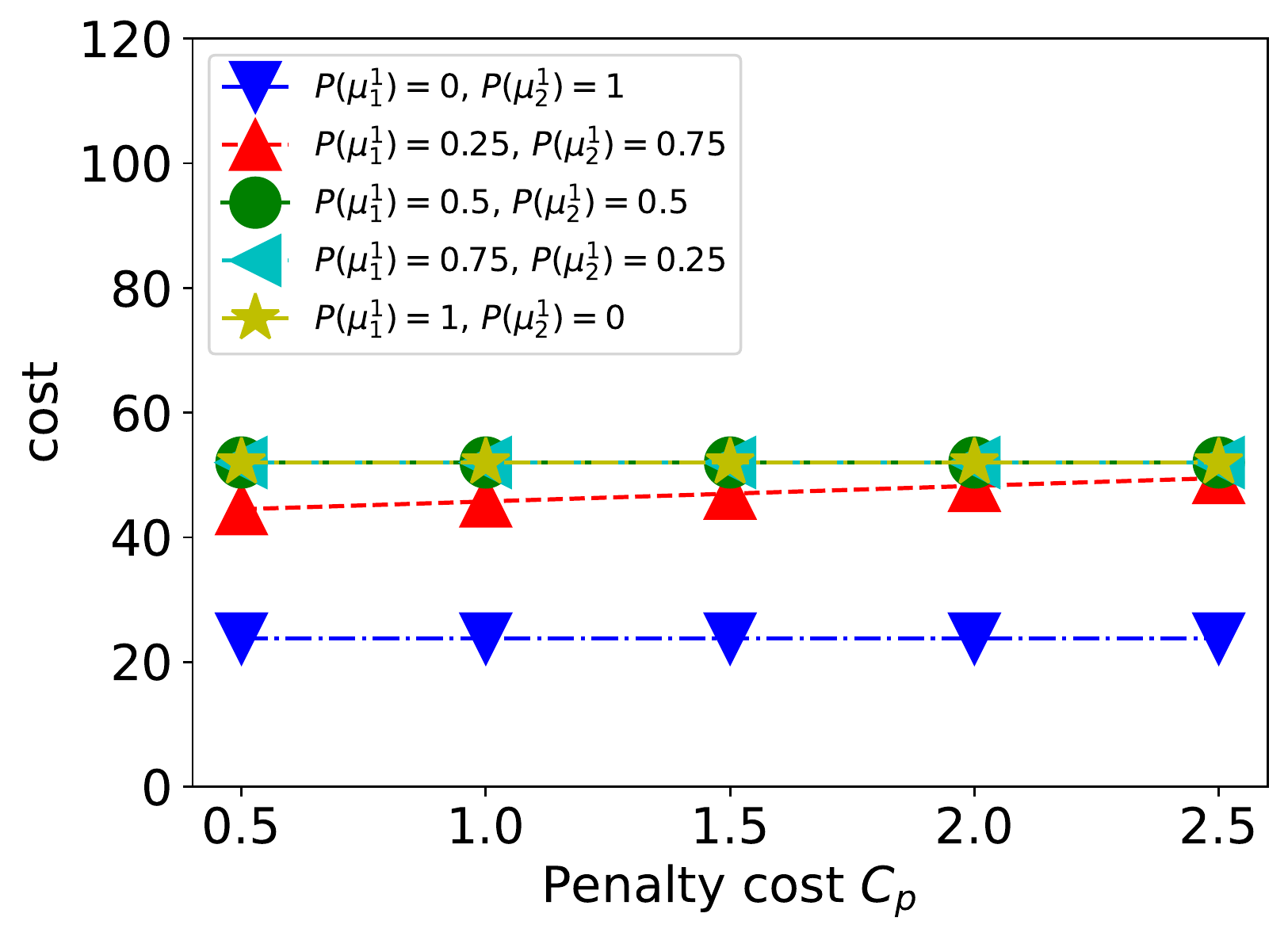}\par
\caption{The cost of the network by varying the penalty cost and the weather probability.}
\label{fig:pro}
\end{multicols}
\end{figure*}

We obtain the finite set $\mathcal{W}$ from the road traffic data set released by Land Transport Authority Singapore~\cite{data}. As depicted in Fig.~\ref{fig:traffic camera}, there are 72 cameras in Singapore to monitor the traffic flow. We generate 1 set of the data in real-time to obtain $\mathcal{Y}$. The data are in the dimension of $row\times column$. Following~\cite{dutta2019optimal} we are only considering square matrices, therefore, $W~\!\!\!\!\!=~\!\!\!\!\!row~\!\!\!\!\!=~\!\!\!\!\!column$. The probability distribution of the demand size is shown in Fig.~\ref{fig:probability dis} and $\mathcal{W}=~\{240, 260, 480, 1080\}$. We use scenarios to represent the weather, demand, and shortfall uncertainties. In each scenario, each uncertainty will have a corresponding value. For example, when $G^{\bar{t}}_y(\mu_i^{\bar{t}})=1$. It means that the weather uncertainty in scenario $i$ in mobile charging station $y$ at time slot $\bar{t}$ is represented by the value of 1. For the road traffic data set, we obtain the dimension of the data, and we post-process the data to obtain $\mathcal{W}$ which is the demand size. We use the demand size to represent the actual demand uncertainty. Similar to weather uncertainty, $\mathcal{W}$ is represented by its corresponding values $240$, $260$, $480$, and $1080$. As shown in Fig.~\ref{fig:insert}, weather, demand, and shortfall uncertainties can be injected into the GAMS script using their corresponding values.

%The complexity for the two-stage and multi-stage SIP is \#P-hard if all the scenarios are considered~\cite{dyer2006computational}. Therefore, 
\subsection{Allocation of UAV}
We first evaluate the UAV type allocation for each mobile charging station. We consider a two-stage SIP, where the first stage is~(\ref{2sip_1}) and the second stage is~(\ref{2sip_2}). In this UAV type allocation, we consider two scenarios of weather condition, i.e., strong wind and weak wind, which are $|\gamma^1| =2$~\cite{chaisiri2011optimization}. The two scenarios are that all locations have strong wind $\mu^1_1$ and no location has strong wind $\mu^1_2$. We consider a stochastic system with $\mathcal{P}(\mu^1_1)=0.3$ and $\mathcal{P}(\mu^1_2)=0.7$. We perform the sensitivity analysis by varying the incurred cost and the weather condition probability.
\subsubsection{Penalty cost $C_p$}\label{penalty} We evaluate the UAV type allocation by varying the penalty cost $C_p$, and the result is presented in Fig.~\ref{fig:penalty}. When the penalty cost is low, the application owner still prefers to reserve the small UAV even though there is a crashing probability. However, the decision changed when the $C_p \geq1.5$. The cost of the correction action, which is the on-demand cost of using a type 3 UAV, becomes much higher. $\bar{A}_3$ has to make a reservation on the type 3 UAV in the initial phase. 
\subsubsection{Probability of the weather condition} Next, we consider the same setting as that in Section~\ref{penalty}. We investigate the impact on the network by varying both the weather probability $P(\mu^1_1)$ and $C_p$. The result is shown in Fig. \ref{fig:pro}. When the weather is good $P(\mu^1_1)=0$, all the charging stations will choose to use a type 1 UAV in stage 1 since there is no strong wind. When the probability of bad weather increases, e.g., $P(\mu^2_1)\geq0.5$ all the charging stations tend to choose type 3 UAVs even when the penalty cost is low due to a high probability of UAV crashes.

\subsection{Allocation of Task}
We now perform the sensitivity analysis on task allocation using the UAV type allocation derived when $C_p=2$. We perform the sensitivity analysis by varying the number of stages $z$, hovering cost $C^{thresh}_{y,x^{\bar{t}}}$, the probability of shortfalls, recovery threshold $k$ as well as the types of UAV use, and we discuss the results in this subsection.
\subsubsection{Cost structure} 
\begin{figure*}
\centering
\begin{multicols}{3}
\includegraphics[width=0.9\columnwidth]{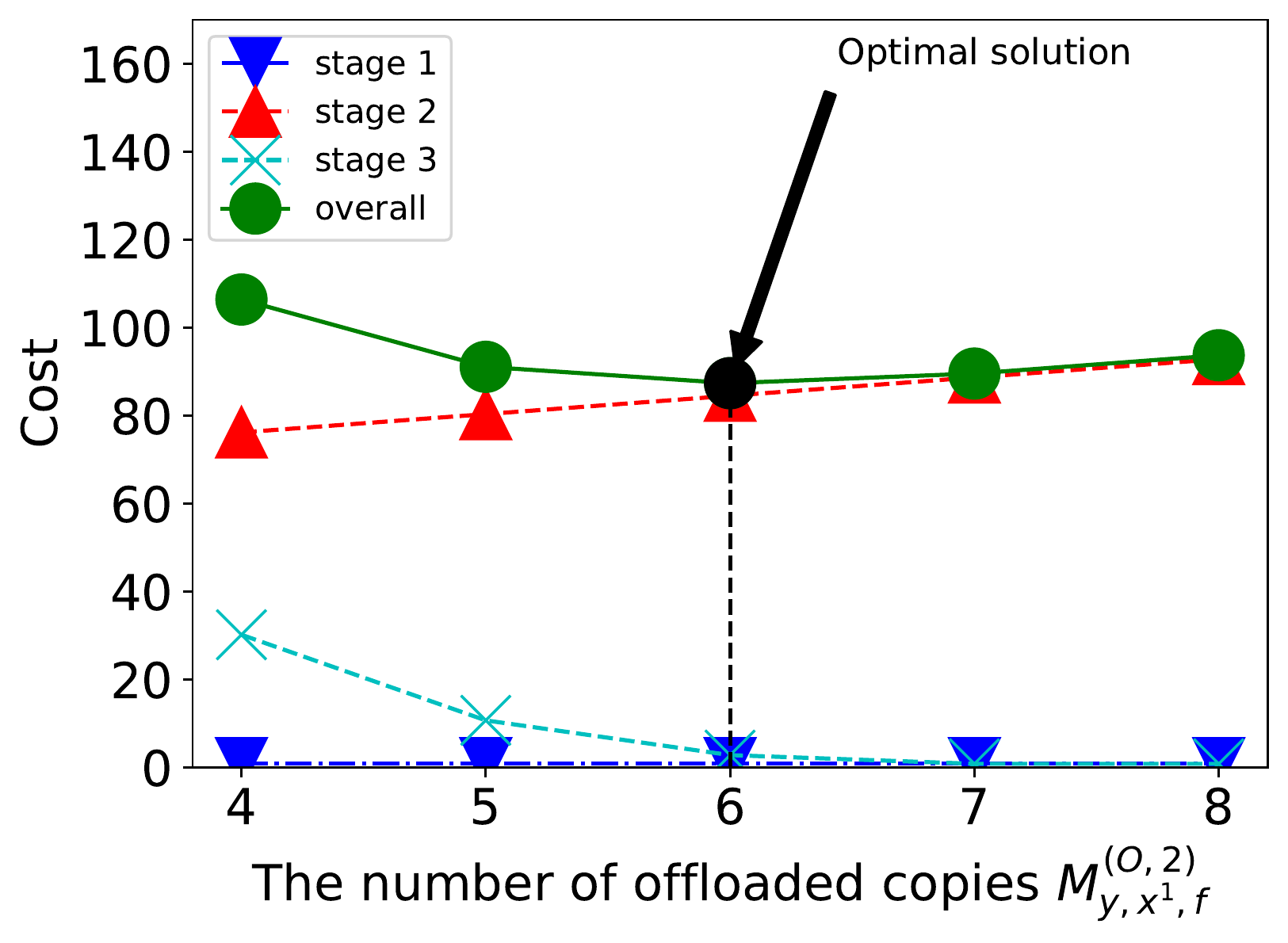}
\caption{The optimal solution in a simple three-stage SIP UAV network.}
\label{fig:optimal cost}
\includegraphics[width=0.9\columnwidth]{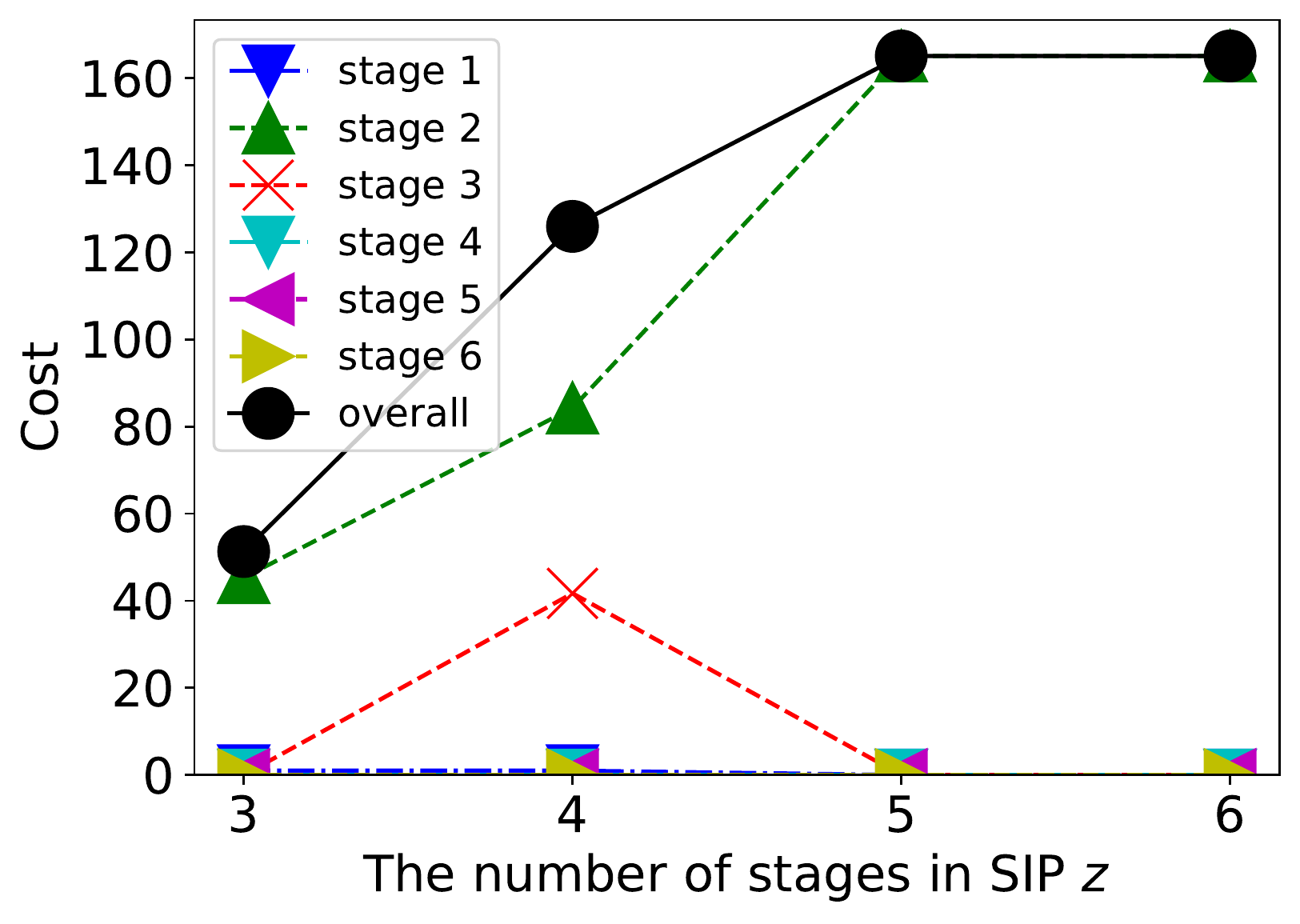}\par
\caption{The cost of the network by varying the number of stages in SIP. Considered only UAV 1, BS 1 and BS 2.}
\label{fig:1UAV}
\includegraphics[width=0.9\columnwidth]{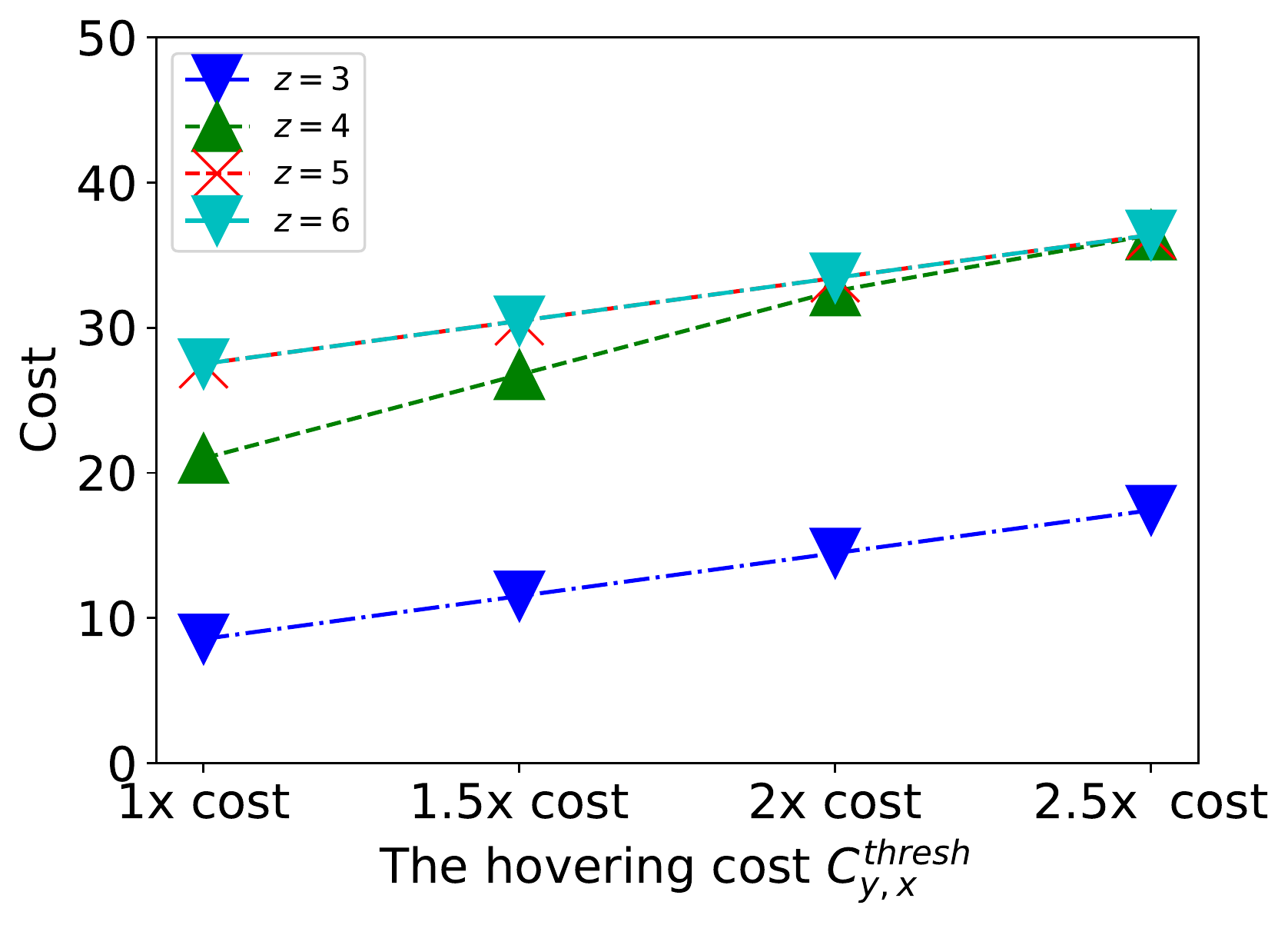}\par
\caption{The cost of the network by varying the hovering cost $C^{thresh}_{y,x^{\bar{t}}}$.}
\label{fig:varyinghovering}
\end{multicols}
\end{figure*}

We first study the cost structure of the network. As an illustration, a primitive UAV network is considered with zero local computation $M^{L,2}_{y,x}=0$, $M^{L,3}_{y,x}=0$, and we consider a three-stage SIP, where the first stage is (\ref{sip_1}), the second stage is (\ref{sip_2}) and the third stage is (\ref{sip_4}). Furthermore, we first consider six mobile charging stations, one BS, one demand scenario $|\Theta^1|=1$ and $|\Omega^{(3,1)})|=1$. The demand is 1080. The shortfall scenario is that there are shortfalls from the copies that the UAV has offloaded. Since $k=4$, the number of copies that are offloaded $M^{(O,2)}_{y,x,f}$ begins with 4. In Fig.~\ref{fig:optimal cost}, the costs in the first, second, third stages and the total cost under the different number of the offloaded copies are presented. Since offloading is the only decision, the edge servers subscription cost in stage one should always paid. In the second stage, the cost is expected to increase as the number of copies that are offloaded increases. However, the cost in the third stage after knowing the actual shortfall decreases as the number of offloaded copies increases since the UAV needs to perform fewer re-computation to match the shortfall. Fig.~\ref{fig:optimal cost} shows the optimal solution in this simple network. It can be identified that even in this simple network, the optimal solution is not trivial to obtain due to the uncertainty of shortfall. For example, the optimal cost is not the point where the cost in the second stage is the lowest of the cost in the third stage is the lowest. Therefore, SIP formulation is required to guarantee the minimum cost to the UAVs.

\begin{figure*}
\centering
\begin{multicols}{3}
\includegraphics[width=0.9\columnwidth]{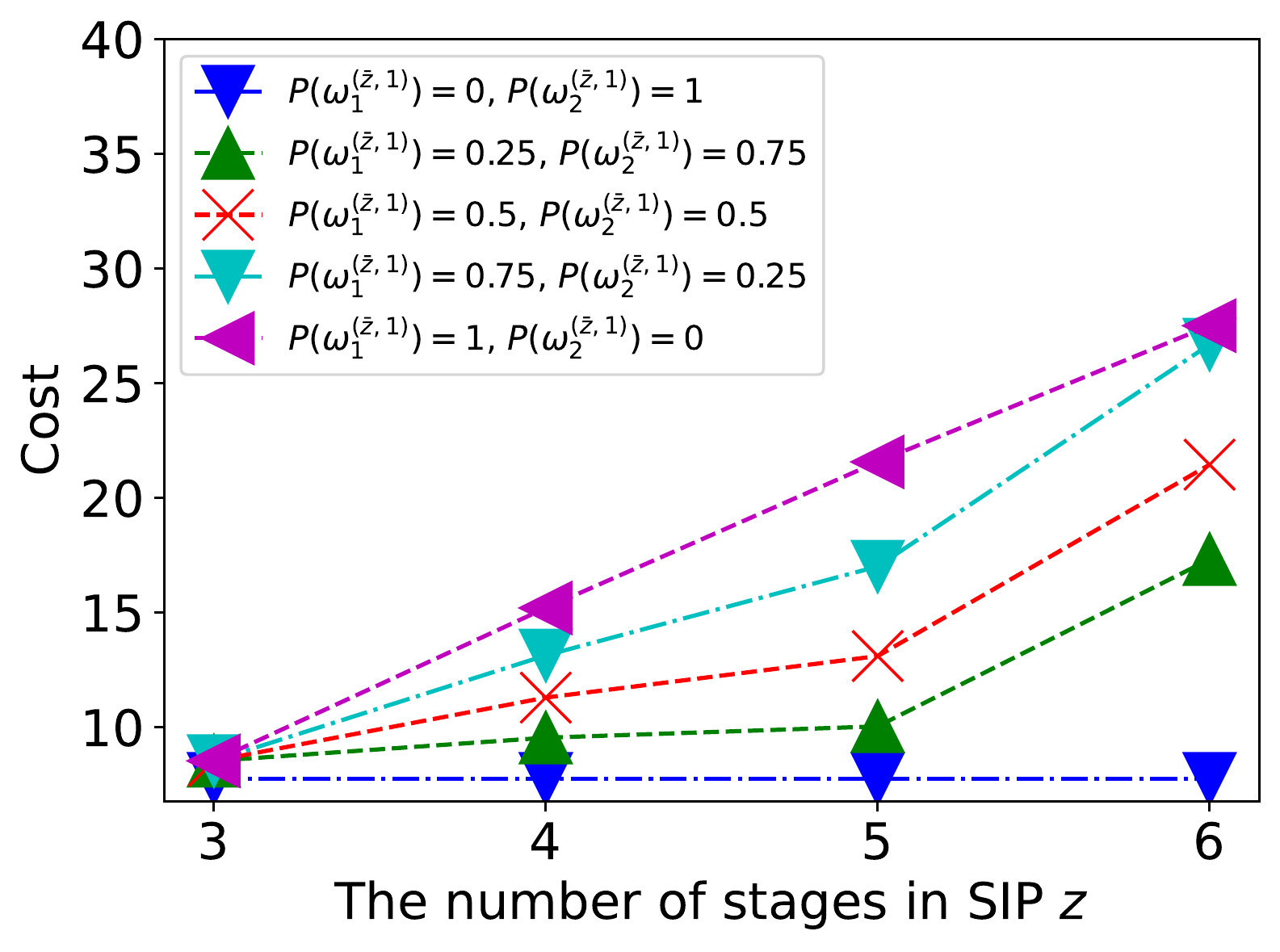}\par
\caption{The cost of the network by varying the probability of shortfall in each stage.}
\label{fig:p}
\includegraphics[width=0.9\columnwidth]{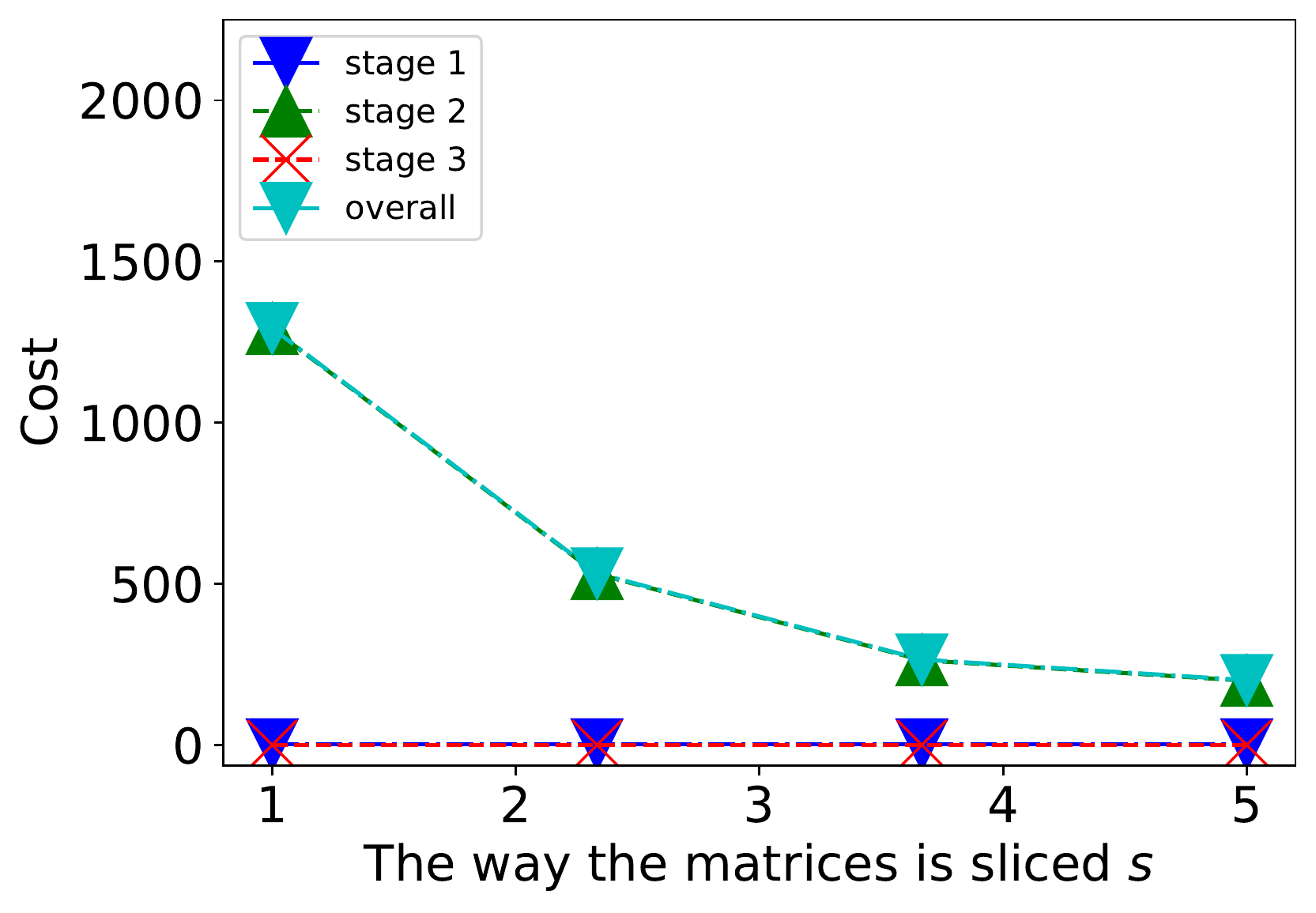}
\caption{The cost of the network by varying the variable $s$ to change the recovery threshold.}
\label{fig:varyingk}
\includegraphics[width=0.9\columnwidth]{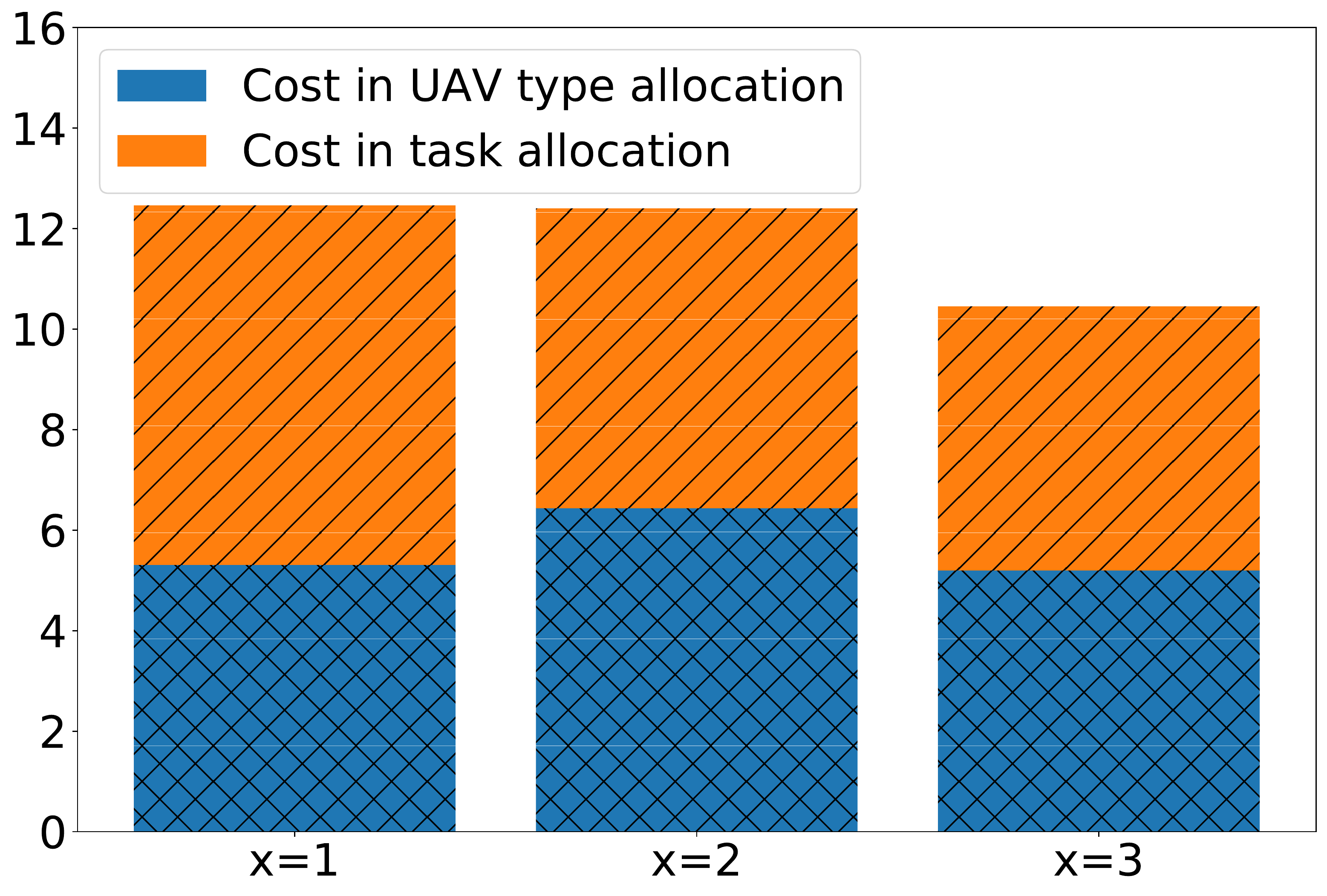}
\caption{The overall network cost when the UAV type $x$ is varied.}
\label{fig:uav type cost}
\end{multicols}
\end{figure*}

\subsubsection{Number of stages $z$}\label{stages}We consider the case with four demand scenario $|\Theta^1|=4$, one shortfall scenario $|\Omega^{(\bar{z},1)}|=1$ for each SIP stage, and there is a shortfall in each stage. The four demand scenarios are made up of 4 demand size each, $\lambda_1^1 = 480$, $\lambda_2^1 = 240$, $\lambda_3^1 = 1080$ and $\lambda_4^1 = 360$. We vary the value of $z$ from 3 to 6. The cost of the network is shown in Fig. \ref{fig:1UAV} and Table \ref{table:variable} indicates the value of the variables. The number of variables in the six-stage SIP is large. Therefore, for illustration, we only indicate the value for all the variables in mobile charging station 1 when $z=3$. Then, We use the result from $z=3$ to explain the findings. When the data size is 240, the offloading cost is more expensive than the local computation cost. Therefore, the UAV will perform local computation in stage 2 when the data size is 240. $M_{1,3^1}^{(L,2)}(\lambda_2^1) =4$, $M_{1,3^1,f}^{(O,2)}(\lambda_2^1) = 0$, $M_{1,3^1,f}^{(L,3)}(\lambda_2^1) = 0$ and $M_{1,3^1,f}^{(O,3)}(\lambda_2^1) = 0$. However, for the rest of the data size, the offloading action is chosen. The UAV will offload more in stage 2 to overcome the shortfall, and this offloading cost is cheaper than the local computation cost. The total number of copies offloaded to the edge servers in BSs 1 and 2 is 24 each. The UAVs will always offload to the edge servers that is closer to it as the cost is lower. To support the offloading process, 24 edge servers have to be placed in both BSs 1 and 2. Since the shortfalls exist in every stage, as the number of stages increases, the UAV will perform more local computation in stage 2. When $z\geq5$, the UAV will only consider local computation regardless of the demand, and the overall cost will remain the same. Since there is no offloading process, edge servers' locations are no longer necessary as they are no longer needed.
\subsubsection{Hovering cost $C^{thresh}_{y,x}$}\label{hover} We consider the case with mobile charging station 1 and BS 1. We vary both the hovering cost and the number of stages $z$ to observe the impact on the network. We keep the rest of the parameters the same as Section \ref{stages}. Fig.~\ref{fig:varyinghovering} shows the cost of the network. It can be identified that when the hovering cost increases, it will also increase the re-offloading cost, therefore increasing the network cost. Since there is a shortfall scenario in every stage, the increase of hovering costs in every stage will affect the decision of the UAV. For example, when stage 4 has 2.5 times hovering cost, the UAV will perform full local computation regardless of the data size. Full local computation is also performed when $z\geq5$.

\subsubsection{Probability of the shortfalls} We consider the setting similar to Section~\ref{hover}. There are two shortfall scenarios in each stage $|\Omega^{(\bar{z},1)}|=2$. The two shortfall scenarios are i) all the UAVs have shortfalls $\omega^{(\bar{z},1)}_1$ in stage $\bar{z}$ and ii) all the UAVs do not have shortfall $\omega^{(\bar{z},1)}_2$. We observe the impact on the network by varying both the shortfall probabilities $P(\omega^{(\bar{z},1)}_1)$, $P(\omega^{(\bar{z},1)}_2)$ and the number of stages $z$. The result is shown in Fig. \ref{fig:p}. Since there is no shortfall $P(\omega^{(\bar{z},1)}_1)=0$, when $z$ increases, the decision made by the UAV remains the same. The UAV can choose the cheapest decision, which is the offloading action. When $P(\omega^{(\bar{z},1)}_1)$ increases, the number of copies that are computed locally and offloaded in the earlier stage also increases to reduce the penalty cost by the shortfalls. However, the number of scenarios increases exponentially with $z$~\cite{GamsGuide2013}, and this is the reason that it leads to a sharp increase in cost from stages 5 to 6. For example, in a 6-stage SIP with two scenarios per stage, this results in a total of $5^6=15,625$ 64 scenarios~\cite{GamsGuide2013}. When $z=6$ and $P(\omega^{(\bar{z},1)}_1)=1$, with the high re-offloading cost, full local computation is perform in the earlier stage.

%Since we consider there are shortfall in every stage, this means that as the stages increase the shortfall increase. This explain the huge cost increase from $z=5$ to $z=6$. when $P(\omega^{\bar{z}}_1)$ increases, the re-offloading action is performed only when the shortfall occurs. To prevent having the penalty of $\Tilde{C}$ it will offload more in $z-1$ stage.The huge increase in the cost of the network when $z=6$ is because the additional hovering cost is increases as the number of stages increase. With the low 
\begin{table}
\centering
 \caption{Decision Variable value}
 \begin{tabular}{||c|c| c| c| c||} 
 \hline
 Variable & \textbf{z=3} & $z=4$ & $z=5$ & $z=6$ \\ [0.5ex] 
 \hline\hline
 $M_1^{(s)}$ & 1 & 1 & 0 & 0 \\ 
 \hline
 $M_2^{(s)}$ & 0 & 0 & 0 & 0 \\
 \hline
 $M_{1,3^1}^{(L,2)}(\lambda_1^1)$ & \textbf{0} & 4 & 4 & 4\\
 \hline
  $M_{1,3^1}^{(L,2)}(\lambda_2^1)$ & \textbf{4} & 4 & 4 & 4\\
 \hline
  $M_{1,3^1}^{(L,2)}(\lambda_3^1)$ & \textbf{0} & 0 & 4 & 4\\
 \hline
  $M_{1,3^1}^{(L,2)}(\lambda_4^1)$ & \textbf{0} & 4 & 4 & 4\\
 \hline
 $M_{1,3^1,1}^{(O,2)}(\lambda_1^1)$ & \textbf{8} & 0 & 0 & 0 \\
 \hline
 $M_{1,3^1,1}^{(O,2)}(\lambda_2^1)$ & \textbf{0} & 0 & 0 & 0 \\
 \hline
 $M_{1,3^1,1}^{(O,2)}(\lambda_3^1)$ & \textbf{8} & 4 & 0 & 0 \\
 \hline
 $M_{1,3^1,1}^{(O,2)}(\lambda_4^1)$ & \textbf{8} & 0 & 0 & 0 \\
 \hline
 $M_{1,3^1}^{(L,3)}(\lambda_1^1,\omega^{(3,1)})$ & 0 & 0 & 0 & 0 \\
 \hline
  $M_{1,3^1}^{(L,3)}(\lambda_2^1,\omega^{(3,1)})$ & 0 & 0 & 0 & 0 \\
 \hline
  $M_{1,3^1}^{(L,3)}(\lambda_3^1,\omega^{(3,1)})$ & 0 & 0 & 0 & 0 \\
 \hline
  $M_{1,3^1}^{(L,3)}(\lambda_4^1,\omega^{(3,1)})$ & 0 & 0 & 0 & 0 \\
 \hline
 $M_{1,3^1,1}^{(O,3)}(\lambda_1^1,\omega^{(3,1)})$ & 0 & 0 & 0 & 0 \\
 \hline
  $M_{1,3^1,1}^{(O,3)}(\lambda_2^1,\omega^{(3,1)})$ & 0 & 0 & 0 & 0 \\
 \hline
  $M_{1,3^1,1}^{(O,3)}(\lambda_3^1,\omega^{(3,1)})$ & 0 & 8 & 0 & 0 \\
 \hline
  $M_{1,3^1,1}^{(O,3)}(\lambda_4^1,\omega^{(3,1)})$ & 0 & 0 & 0 & 0 \\
 \hline
  Total number of edge server in BS 1 & \textbf{24} & 24 & 0 & 0 \\
 \hline
   Total number of edge server in BS 2 & \textbf{24} & 24 & 0 & 0 \\
 \hline
\end{tabular}
\label{table:variable}
\end{table}
\subsubsection{Recovery threshold k}\label{reco} Similar to Section~\ref{hover}, we consider three-stage SIP with four demand scenarios $|\Theta^1|=4$, one shortfall scenario $|\omega^{(3,1)}|=1$ and there is shortfalls in each offloading process. We monitor the impact of the network by varying the recovery threshold. We set $m=4$ and we vary recovery threshold $k$ by changing $s$ from 1 to 5. Since $k$ is controlled by (\ref{1_equ}), as $s$ increases, $k$ varies from 16 to 6. When $s=1$, the recovery threshold is the highest $k=16$. The result is shown in Fig. \ref{fig:varyingk}. The stage 1 cost remains constant as the BS is always subscribed. The cost of the network is also the highest as the UAV has to compute the most number of copies. When the data size is 240, 360, and 480, in stage 2, the UAV prefers to compute most of the copies locally as the local computation cost is much lower than the offloading cost plus the re-offloading cost in stage 3. The reason is that when $k$ is high, the number of shortfalls is also high, which increases the re-offloading cost. This is why the stage 3 cost is zero. As $s$ increases, the value of $k$ decreases, the cost in local computation increases. Therefore, instead of local computation, some of the UAVs perform offloading process in stage 2. Throughout the simulations, the cost in stage 3 is zero. With the additional penalty and hovering cost, no UAVs prefer to compute the shortfalls in stage 3.
%Fourthly, we evaluate the effect of the number of UAVs in the SIP network with $z=3$. The network is initialized with only $UAV_1$ and $n_1,n_2 = 30$. The total cost of the network is monitored as the number of UAVs joining the network increases. Figure~\ref{fig:effectUAV} displays the result from this set-up. When the number of UAVs in the network is more than 8, we can observe a sharp increase in this network's cost far beyond the preceding trend, as indicated by the dotted line. The total number of workers in each BS that can work on the matrix multiplication is $n_f$. In this case, all the workers from the BSs are fully occupied when the number of UAVs exceeded 8. Instead of offloading to the BSs, the rest of the UAVs have to compute the copy locally, which is more expensive and leads to a sharp cost increase.

\subsubsection{UAV types}Similar to Section~VI-A1 and Section~VI-B5, we consider 1 UAV and $C_p=2$ and investigate the differences in cost when we change the UAV type. The overall network cost is shown in Fig.~\ref{fig:uav type cost}. If the service provider initially uses type 1 and 2 UAVs, and when the wind is strong, the UAV cannot have sufficient energy to withstand the strong wind (weather uncertainty). As a result, the service provider has to deploy another UAV (type 3) to perform the task. However, if the service provider initially deploys a type 3 UAV, it will have sufficient energy to withstand the strong wind. There is no additional correction cost. Therefore, the UAV type allocation cost is cheaper if the service provider chooses to book type 3 UAV in advance. Furthermore, using type 3 UAV in task allocation makes the computation capability stronger than the other UAV types. As a result, it will have a shorter latency and will lead to a lower computation cost. Hence, the computation cost is also the lowest when a type 3 UAV is used. Therefore, the overall network cost is the lowest when a type 3 UAV is used.
\begin{figure}
\centering
\includegraphics[width=0.9\columnwidth]{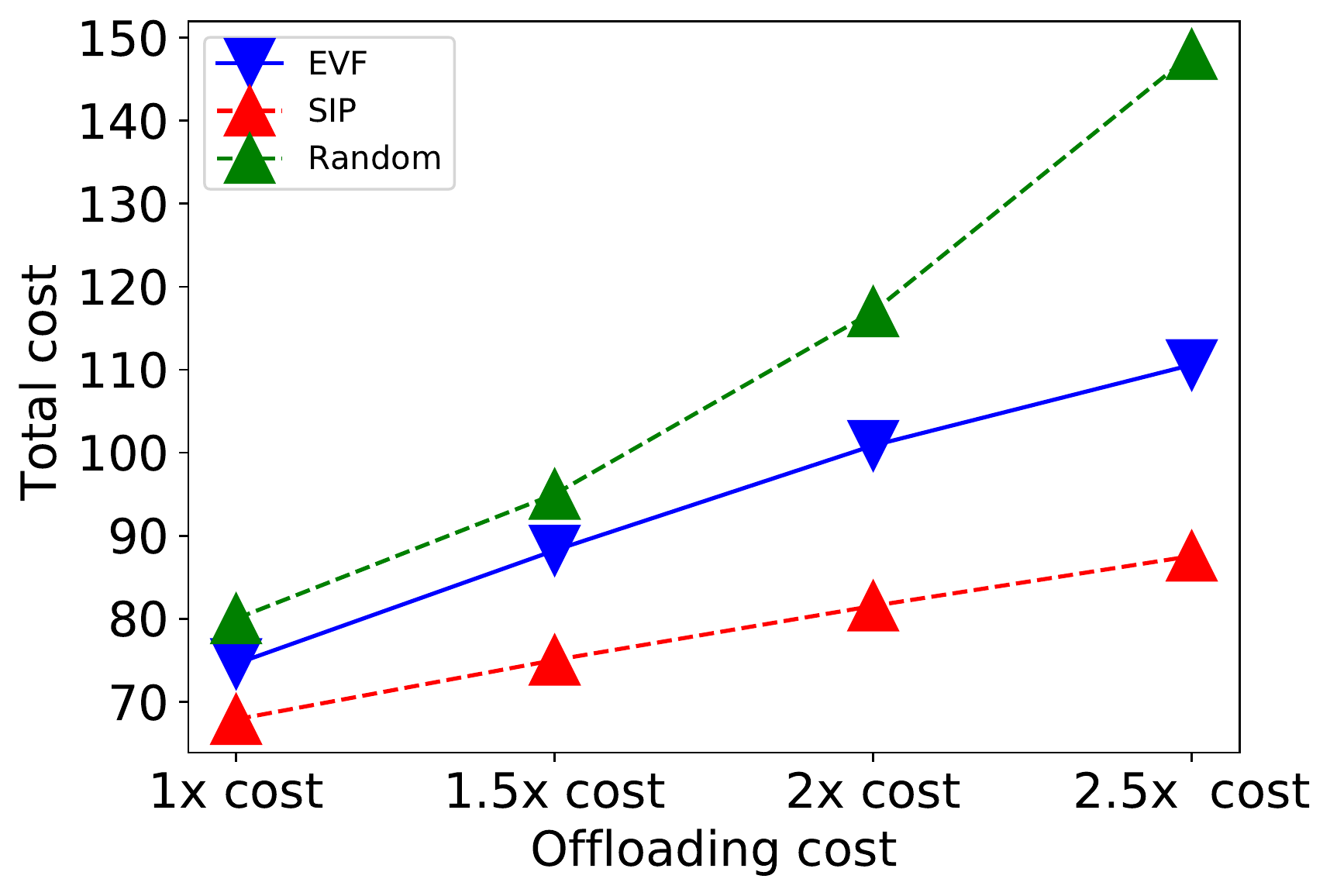}\par
\caption{SIP comparing with EVF and random scheme.}
\label{fig:evf}
\end{figure}
\subsection{Comparing between EVF, SIP and random scheme} We set $z=3$ and compare the SIP with Expected-Value Formulation (EVF)~\cite{5394134} as well as the random scheme. Expected-value formulation uses the average values of shortfall as well as the demand and solves a DIP. For EVF, the total number of copies ($M_{y,x^{\bar{t}}}^{(L)}+M_{y,x^{\bar{t}},f}^{(O)}$) is fixed using the average value of shortfall and demand, an approximation scheme. In a random scheme, the values of the decision variables are randomly generated. We vary the price of the offloading action to compare the difference between EVF, SIP, and random schemes. This cost is combined with the UAV type allocation cost to observe the total cost of the network. Fig.~\ref{fig:evf} depicts the comparison result. As shown in the result, EVF and random scheme cannot adapt to the change in cost. On the other hand, SIP can always achieve the best solution among the three to reduce the shortfall cost.

\section{Conclusion}\label{conclusion}
In this paper, we have proposed the Stochastic Coded Offloading Scheme (SCOS) that employs a CDC technique to combines with computation offloading to minimize the energy consumption of the UAV network. We have first formulated the SCOS as the two-stage stochastic programming to optimize the UAV type allocation to account for the weather uncertainty. Then, we have formulated the $z$-stage stochastic programming to account for data size and task completion uncertainty. We have also conducted numerical experiments to verify that our proposed SCOS can optimize the total cost and the UAVs’ energy consumption. Compared to the benchmark, SCOS based on SIP can achieve the best solution as it can adapt to changes in data size and task failure probability. We also know that a three-stage SIP is enough to formulate the problem through the simulations as it has the lowest cost.

\section*{Acknowledgement}
This research is supported, in part, the programme DesCartes and is supported by the National Research Foundation, Prime Minister’s Office, Singapore under its Campus for Research Excellence and Technological Enterprise (CREATE) programme, Alibaba Group through Alibaba Innovative Research (AIR) Program and Alibaba-NTU Singapore Joint Research Institute (JRI), the National Research Foundation, Singapore under the AI Singapore Programme (AISG) (AISG2-RP-2020-019), WASP/NTU grant M4082187 (4080), Singapore Ministry of Education (MOE) Tier 1 (RG16/20), in part by the SUTD SRG-ISTD-2021-165, the SUTD-ZJU IDEA Grant (SUTD-ZJU (VP) 202102), and the SUTD-ZJU IDEA Seed Grant (SUTD-ZJU (SD) 202101). This research is also supported by, in part, the NSF CNS-2107216 and EARS-1839818; in part by the National Research Foundation of Korea (NRF) Grant funded by the Korean Government (MSIT) under Grant 2021R1A2C2007638 and the MSIT under the ICT Creative Consilience program (IITP-2020-0-01821) supervised by the IITP.
% or


\appendices

\section{Constraints of $z$-stage SIP System Model}\label{append}
\begin{align}
    \hspace*{0mm}&\sum_{y\in\mathcal{Y}}M_{y,x^{\bar{t}},f}^{(O,2)}(\lambda_i^{\bar{t}}) \leq \sigma M_f^{(s)},  &\hspace*{+0mm} \forall\bar{t}\in\mathcal{T},\forall f \in\mathcal{F},\forall \lambda_i^{\bar{t}}\in\Theta^{\bar{t}},\label{cons1}
\end{align}
\begin{align}
    \vdots\hspace*{0mm}\nonumber\noindent
\end{align}
\begin{align}
    \sum_{y\in\mathcal{Y}}M_{y,x^{\bar{t}},f}^{(O,\hat{z})}(\lambda_i^{\bar{t}},\ldots,\omega_i^{(\bar{z},\bar{t})}) \leq \sigma M_f^{(s)},\hspace*{+27mm}\nonumber\\
    \forall\bar{t}\in\mathcal{T},\forall f \in\mathcal{F}, \forall \lambda_i^{\bar{t}}\in\Theta^{\bar{t}},\forall\omega_i^{(3,\bar{t})}\in\Omega^{(3,\bar{t})},\ldots,\nonumber\\ \forall\omega_i^{(\bar{z},\bar{t})}\in\Omega^{(\bar{z},\bar{t})},\label{cons2}\hspace*{-0mm}
\end{align}
\begin{align}
    M_{y,x^{\bar{t}},f}^{(O,3)}(\lambda_i^{\bar{t}},\omega_i^{(3,\bar{t})}) \leq \sigma M^{(TH,3)}_{y,x^{\bar{t}},f}(\lambda_i^{\bar{t}},\omega_i^{(3,\bar{t})}), \hspace*{+10mm}\nonumber\\ \forall\bar{t}\in\mathcal{T},\forall y\in \mathcal{Y},\forall f \in\mathcal{F},\forall \lambda_i^{\bar{t}}\in\Theta^{\bar{t}},\forall\omega_i^{(3,\bar{t})}\in\Omega^{(3,\bar{t})},\label{cons20}
\end{align}
\begin{align}
    \vdots\hspace*{0mm}\nonumber\noindent
\end{align}
\begin{align}
    M_{y,x^{\bar{t}},f}^{(O,\hat{z})}(\lambda_i^{\bar{t}},\ldots,\omega_i^{(\bar{z},\bar{t})}) \leq \sigma M^{(TH,\bar{z})}_{y,x^{\bar{t}},f}(\lambda_i^{\bar{t}},\ldots,\omega_i^{(\bar{z},\bar{t})}), \hspace*{+0mm}\nonumber\\\forall\bar{t}\in\mathcal{T},\forall y\in \mathcal{Y},\forall f \in\mathcal{F},\forall \lambda_i^{\bar{t}}\in\Theta^{\bar{t}},\forall\omega_i^{(3,\bar{t})}\in\Omega^{(3,\bar{t})},\ldots,\nonumber\\ \forall\omega_i^{(\bar{z},\bar{t})}\in\Omega^{(\bar{z},\bar{t})},\label{cons21}
\end{align}
\begin{align}
    \sum_{f\in\mathcal{F}}M_{y,x^{\bar{t}},f}^{(O,3)}(\lambda_i^{\bar{t}},\omega_i^{(3,\bar{t})}) \geq \mathbb{F}^{(3,\bar{t})}_y(\lambda_i^{\bar{t}},\omega_i^{(3,\bar{t})})\hspace*{+20mm} \nonumber\\\biggl(\bar{A}^{(3,\bar{t})}_y(\lambda_i^{\bar{t}},\omega_i^{(3,\bar{t})})-M^{(L,3)}_{y,x^{\bar{t}}}(\lambda_i^{\bar{t}},\omega_i^{(3,\bar{t})})\biggr), \hspace*{+0mm}\nonumber\\\forall\bar{t}\in\mathcal{T},\forall y\in \mathcal{Y},\forall \lambda_i^{\bar{t}}\in\Theta^{\bar{t}}, \forall\omega_i^{(3,\bar{t})}\in\Omega^{(3,\bar{t})},\hspace*{-0mm}\label{cons5}
\end{align}
\begin{align}
    \vdots\hspace*{0mm}\nonumber\noindent
\end{align}
\begin{align}
    \sum_{f\in\mathcal{F}}M_{y,x^{\bar{t}},f}^{(O,\hat{z})}(\lambda_i^{\bar{t}},\ldots,\omega_i^{(\bar{z},\bar{t})}) \geq \biggl(\bar{A}^{(\bar{z},\bar{t})}_y(\lambda_i^{\bar{t}},\ldots,\omega_i^{(\bar{z},\bar{t})})-\hspace*{+6mm}\nonumber\noindent\\ M^{(L,\hat{z})}_{y,x^{\bar{t}}}(\lambda_i^{\bar{t}},\ldots,\omega_i^{(\bar{z},\bar{t})})\biggr) \mathbb{F}^3_y(\lambda_i^{\bar{t}},\omega_i^{(3,\bar{t})})\ldots\mathbb{F}^{\bar{z}}_y(\lambda_i^{\bar{t}},\ldots,\omega_i^{(\bar{z},\bar{t})}), \hspace*{+0mm}\nonumber\noindent\\ \forall\bar{t}\in\mathcal{T},\forall y\in \mathcal{Y},\forall \lambda_i^{\bar{t}}\in\Theta^{\bar{t}}, \forall\omega_i^{(3,\bar{t})}\in\Omega^{(3,\bar{t})},\ldots,\nonumber\noindent\\ \forall\omega_i^{(\bar{z},\bar{t})}\in\Omega^{(\bar{z},\bar{t})},\label{cons6}
\end{align}
\begin{align}
    M_{y,x^{\bar{t}}}^{(L,2)}(\lambda_i^{\bar{t}}) + M_{y,x^{\bar{t}},f}^{(O,2)}(\lambda_i^{\bar{t}}) \geq k, \hspace*{+32mm} \nonumber\\\forall\bar{t}\in\mathcal{T}, \forall y\in \mathcal{Y}, \forall f\in \mathcal{F}, \forall \lambda_i^{\bar{t}}\in\Theta^{\bar{t}},\hspace*{-0mm}\label{cons7}
\end{align}
\begin{align}
    M_{y,x^{\bar{t}}}^{(L,2)}(\lambda_i^{\bar{t}}) + \sum_{f\in\mathcal{F}}M_{y,x^{\bar{t}},f}^{(O,2)}(\lambda_i^{\bar{t}}) + \mathbb{F}_y^{(3,\bar{t})}(\lambda_i^{\bar{t}},\omega_i^{(3,\bar{t})})\hspace*{+15mm} \nonumber\\M_{y,x^{\bar{t}}}^{(L,3)}(\lambda_i^{\bar{t}},\omega_i^{(3,\bar{t})}) +  \sum_{f\in\mathcal{F}}\mathbb{F}^{(3,\bar{t})}_y(\lambda_i^{\bar{t}},\omega_i^{(3,\bar{t})})M_{y,x^{\bar{t}},f}^{(O,3)}(\lambda_i^{\bar{t}},\omega_i^{(3,\bar{t})})  \hspace*{+0mm}\nonumber\\- \mathbb{F}^{(3,\bar{t})}_y(\lambda_i^{\bar{t}},\omega_i^{(3,\bar{t})})\biggl(\bar{A}^{(3,\bar{t})}_y(\lambda_i^{\bar{t}},\omega_i^{(3,\bar{t})})-M^{(L,2)}_{y,x^{\bar{t}}}(\lambda_i^{\bar{t}},\omega_i^{(3,\bar{t})})\biggr) \nonumber\\\geq k,\hspace*{+0mm}\nonumber\\\forall\bar{t}\in\mathcal{T},
    \forall y\in \mathcal{Y}, \forall f\in \mathcal{F},
     \forall\lambda_i^{\bar{t}}\in\Theta^{\bar{t}}, \forall\omega_i^{(3,\bar{t})}\in\Omega^{(3,\bar{t})},\label{cons8}
\end{align}
\begin{align}
    \vdots\hspace*{0mm}\nonumber\noindent
\end{align}
\begin{align}
    M_{y,x^{\bar{t}}}^{(L,2)}(\lambda^{\bar{t}}) + \sum_{f\in\mathcal{F}}M_{y,x^{\bar{t}},f}^{(O,2)}(\lambda_i^{\bar{t}}) + \ldots + \sum_{f\in\mathcal{F}}\mathbb{F}^{(3,\bar{t})}_y(\lambda_i^{\bar{t}},\omega_i^{(3,\bar{t})})\nonumber\noindent\\
    \ldots\mathbb{F}^{(\bar{z},\bar{t})}_y(\lambda_i^{\bar{t}},\ldots,\omega_i^{(\bar{z},\bar{t})})M_{y,x^{\bar{t}},f}^{(O,\hat{z})}(\lambda_i^{\bar{t}},\ldots,\omega_i^{(\bar{z},\bar{t})}) +
    \nonumber\noindent\\ \mathbb{F}^{(3,\bar{t})}_y(\lambda_i^{\bar{t}},\omega_i^{(3,\bar{t})})\ldots\mathbb{F}^{\bar{z}}_y(\lambda_i^{\bar{t}},\ldots,\omega_i^{(\bar{z},\bar{t})})M_{y,x^{\bar{t}}}^{(L,\hat{z})}(\lambda_i^{\bar{t}},\ldots,\omega_i^{(\bar{z},\bar{t})})  \nonumber\noindent\\- \mathbb{F}^{(3,\bar{t})}_y(\lambda_i^{\bar{t}},\omega_i^{(3,\bar{t})}) \biggl(\bar{A}^{(3,\bar{t})}_y(\lambda_i^{\bar{t}},\omega_i^{(3,\bar{t})})-M^{(L,2)}_{y,x^{\bar{t}}}(\lambda_i^{\bar{t}},\omega_i^{(3,\bar{t})})\biggr)\nonumber\noindent\\  - \ldots - \mathbb{F}^{(3,\bar{t})}_y(\lambda_i^{\bar{t}},\omega_i^{(3,\bar{t})})\ldots \mathbb{F}^{(\bar{z},\bar{t})}_y(\lambda_i^{\bar{t}},\ldots,\omega_i^{(\bar{z},\bar{t})})\nonumber\noindent\\\biggl(\bar{A}^{(\bar{z},\bar{t})}_y(\lambda_i^{\bar{t}},\ldots,\omega_i^{(\bar{z},\bar{t})})-M^{(L,\hat{z})}_{y,x^{\bar{t}}}(\lambda_i^{\bar{t}},\ldots,\omega_i^{(\bar{z},\bar{t})})\biggr) \geq k,\nonumber\noindent\\
    \forall\bar{t}\in\mathcal{T},
    \forall y\in \mathcal{Y}, \forall f\in \mathcal{F},
    \forall\lambda_i^{\bar{t}}\in\Theta^{\bar{t}},
    \forall\omega_i^{(3,\bar{t})}\in\Omega^{(3,\bar{t})}, \ldots,\nonumber\noindent\\
    \forall\omega_i^{(\bar{z},\bar{t})}\in\Omega^{(\bar{z},\bar{t})},\label{cons9}
\end{align}
\begin{align}
    \sum_{y\in\mathcal{Y}}M_{y,x^{\bar{t}},f}^{(O,2)}(\lambda_i^{\bar{t}}) \leq q_f, \hspace*{+0mm}\forall\bar{t}\in\mathcal{T},\forall f \in\mathcal{F},\forall \lambda_i^{\bar{t}}\in\Theta^{\bar{t}},\label{cons10}
\end{align}
\begin{align}
    \vdots\hspace*{0mm}\nonumber\noindent
\end{align}
\begin{align}
    \sum_{y\in\mathcal{Y}} M_{y,x^{\bar{t}},f}^{(O,\hat{z})}(\lambda_i^{\bar{t}},\ldots,\omega_i^{(\bar{z},\bar{t})}) \leq q_f,\hspace*{+25mm} \nonumber\noindent\\\forall\bar{t}\in\mathcal{T}, \forall f \in\mathcal{F},\forall \lambda_i^{\bar{t}}\in\Theta^{\bar{t}}, \forall\omega_i^{(3,\bar{t})}\in\Omega^{(3,\bar{t})}, \ldots,\nonumber\noindent\\
    \forall\omega_i^{(\bar{z},\bar{t})}\in\Omega^{(\bar{z},\bar{t})},\hspace*{-0mm}\label{cons11}
\end{align}
\begin{align}
    M_f^{(s)},M^{(TH,3)}_{y,x^{\bar{t}},f}(\lambda_i^{\bar{t}},\omega_i^{(3,\bar{t})}),\ldots,M^{(TH,\bar{z})}_{y,x^{\bar{t}},f}(\lambda_i^{\bar{t}},\ldots,\omega_i^{(\bar{z},\bar{t})})\nonumber\noindent\\  \in\{0,1\},\hspace*{+0mm}\nonumber\noindent\\ 
    \forall\bar{t}\in\mathcal{T},\forall y\in \mathcal{Y}, \forall f \in\mathcal{F},\forall \lambda_i^{\bar{t}}\in\Theta^{\bar{t}}, \nonumber\noindent\\\forall\omega_i^{(3,\bar{t})}\in\Omega^{(3,\bar{t})}, \ldots,
    \forall\omega_i^{(\bar{z},\bar{t})}\in\Omega^{(\bar{z},\bar{t})}, \hspace*{+0mm}\label{cons12}
\end{align}
\begin{align}
    M_{y,x^{\bar{t}},f}^{(O,2)}(\lambda_i^{\bar{t}}), M_{y,x^{\bar{t}}}^{(L,2)}(\lambda_i^{\bar{t}}), \ldots,M_{y,x^{\bar{t}},f}^{(O,\hat{z})}(\lambda_i^{\bar{t}},\ldots,\omega_i^{(\bar{z},\bar{t})}),\hspace*{+0mm}\nonumber\noindent\\M_{y,x^{\bar{t}}}^{(L,\hat{z})}(\lambda_i^{\bar{t}},\ldots,\omega_i^{(\bar{z},\bar{t})}) \in\{0,1,\ldots\},\nonumber\noindent\\
     \forall\bar{t}\in\mathcal{T},\forall y\in \mathcal{Y}, \forall f \in\mathcal{F},\forall \lambda_i^{\bar{t}}\in\Theta^{\bar{t}}, \nonumber\noindent\\\forall\omega_i^{(3,\bar{t})}\in\Omega^{(3,\bar{t})}, \ldots,
    \forall\omega_i^{(\bar{z},\bar{t})}\in\Omega^{(\bar{z},\bar{t})}.\label{cons13}
\end{align}


%\section{}

%\section*{Acknowledgment}


\ifCLASSOPTIONcaptionsoff
  \newpage
\fi

\bibliographystyle{IEEEtran}
%\typeout{}
\bibliography{mybibliography.bib}


%\vfill

\begin{IEEEbiography}[{\includegraphics[width=1in,height=1.25in,clip,keepaspectratio]{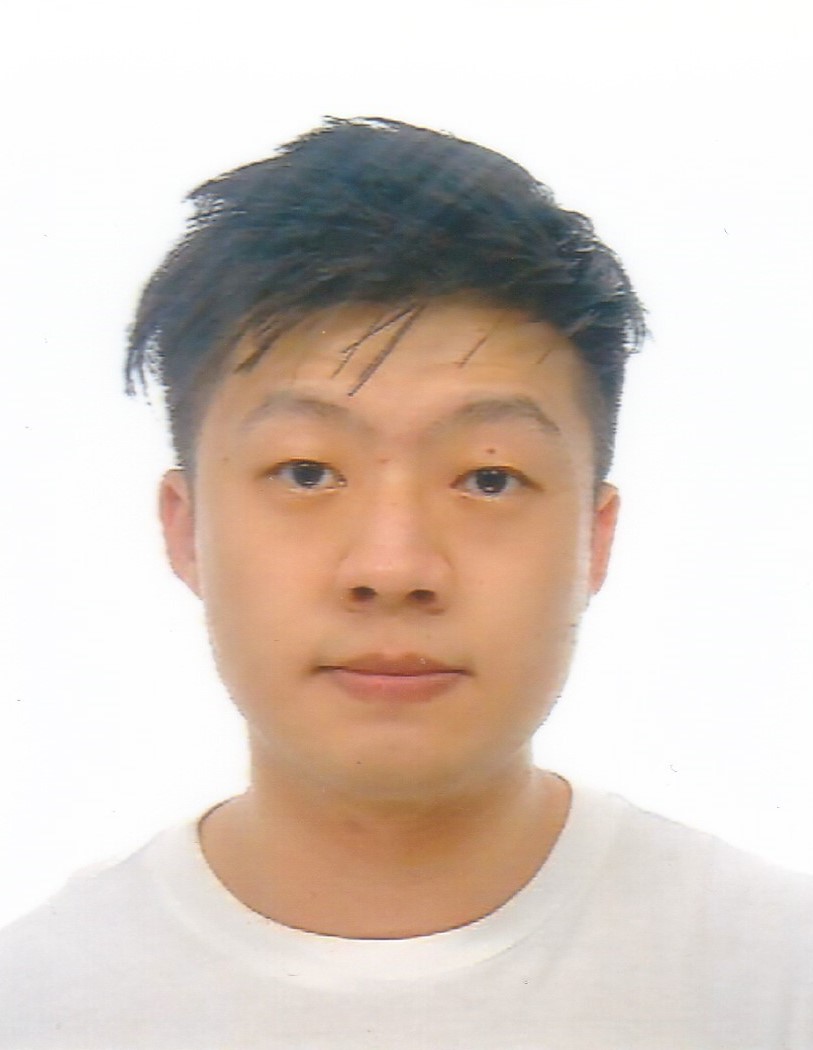}}]{Wei Chong Ng} received B.Eng. degree in electrical and electronic engineering (Highest Distinction) from Nanyang Technological University, Singapore in 2020. He is currently pursuing the Ph.D. degree with Alibaba Group and Alibaba-NTU Joint Research Institute, Nanyang Technological University, Singapore. His research interests include the Metaverse, stochastic integer programming, and edge computing.
\end{IEEEbiography}

\begin{IEEEbiography}[{\includegraphics[width=1in,height=1.25in,clip,keepaspectratio]{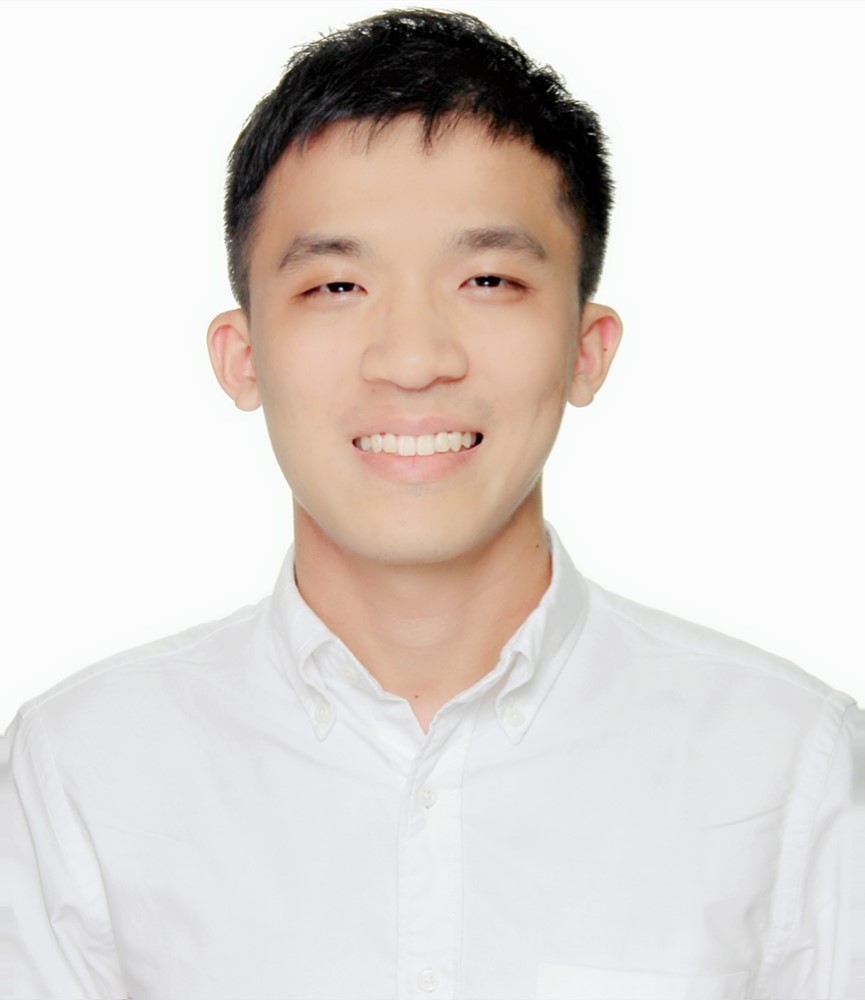}}]{Wei Yang Bryan Lim} is currently an Alibaba Talent Programme PhD candidate with the Alibaba-NTU Joint Research Institute (JRI), Nanyang Technological University (NTU), Singapore. Prior to joining the JRI in 2019, he graduated with two First-Class Honors in Economics and Business Administration (Finance) from the National University of Singapore (NUS). During his PhD candidature, he has won 5 Best Paper Awards including in the IEEE Wireless Communications and Networking Conference (WCNC) and the  IEEE ComSoc SPCC Best Paper Award. He regularly serves as a reviewer in leading journals and flagship conferences and is currently a review board member of the IEEE Transactions on Parallel and Distributed Systems and the assistant to the Editor-in-Chief of the IEEE Communications Surveys \& Tutorials.
\end{IEEEbiography}

\begin{IEEEbiography}[{\includegraphics[width=1in,height=1.25in,clip,keepaspectratio]{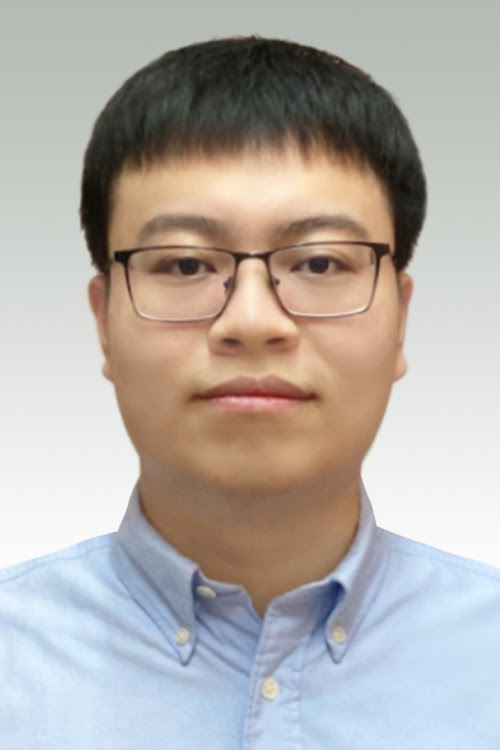}}]{Zehui Xiong} is currently an Assistant Professor in the Pillar of Information Systems Technology and Design, Singapore University of Technology and Design. Prior to that, he was a researcher with Alibaba-NTU Joint Research Institute, Singapore. He received the PhD degree in Nanyang Technological University, Singapore. He was the visiting scholar at Princeton University and University of Waterloo. His research interests include wireless communications, network games and economics, blockchain, and edge intelligence. He has published more than 140 research papers in leading journals and flagship conferences and many of them are ESI Highly Cited Papers. He has won over 10 Best Paper Awards in international conferences and is listed in the World's Top 2\% Scientists identified by Stanford University. He is now serving as the editor or guest editor for many leading journals including IEEE JSAC, TVT, IoTJ, TCCN, TNSE, ISJ. He is the recipient of IEEE TCSC Early Career Researcher Award for Excellence in Scalable Computing, IEEE CSIM Technical Committee Best Journal Paper Award, IEEE SPCC Technical Committee Best Paper Award, IEEE VTS Singapore Best Paper Award, Chinese Government Award for Outstanding Students Abroad, and NTU SCSE Best PhD Thesis Runner-Up Award. He is the Founding Vice Chair of Special Interest Group on Wireless Blockchain Networks in IEEE Cognitive Networks Technical Committee.
\end{IEEEbiography}

\begin{IEEEbiography}[{\includegraphics[width=1in,height=1.25in,clip,keepaspectratio]{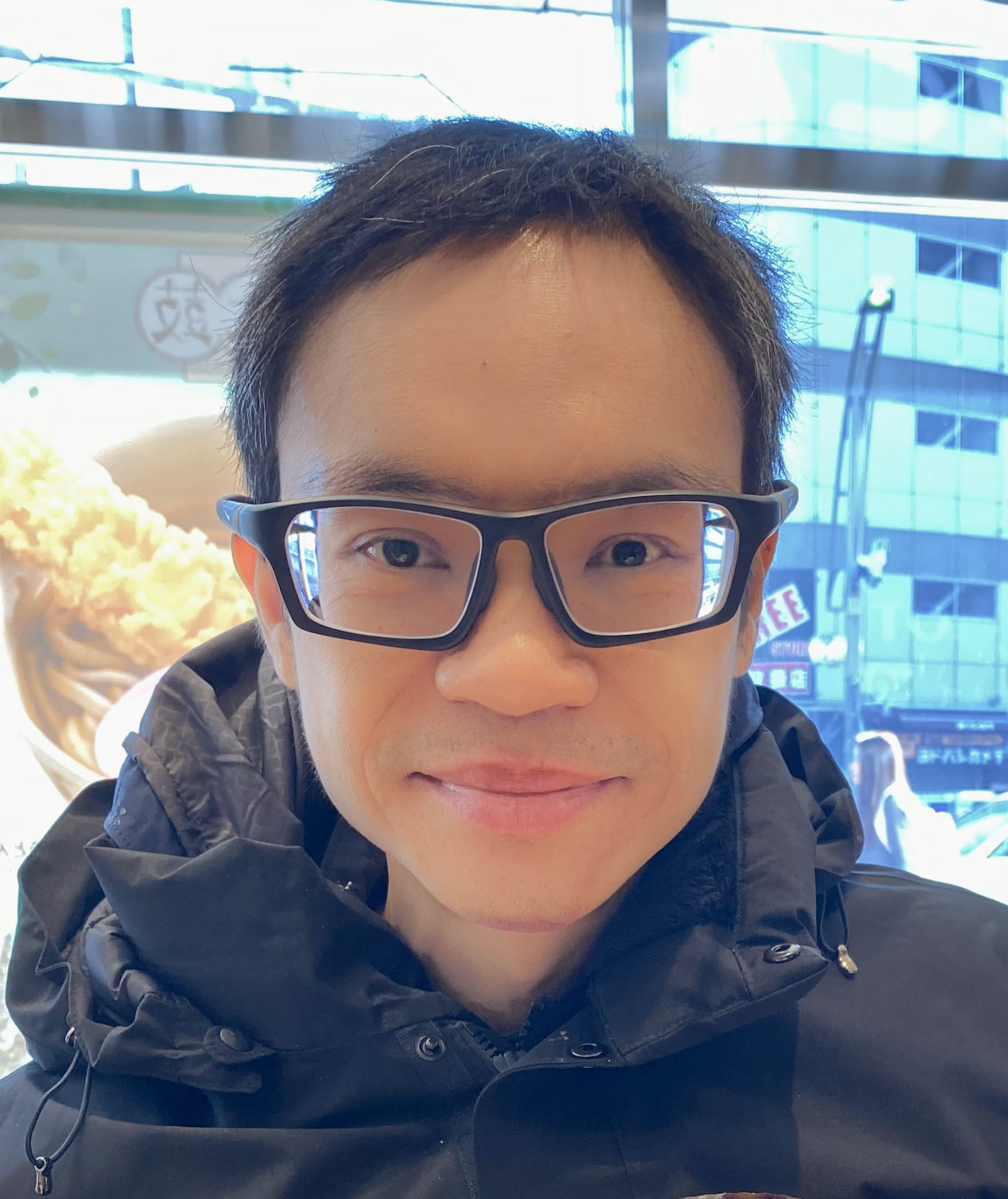}}]{Dusit Niyato} (M'09-SM'15-F'17) is currently a professor in the School of Computer Science and Engineering, at Nanyang Technological University, Singapore. He received B.Eng. from King Mongkuts Institute of Technology Ladkrabang (KMITL), Thailand in 1999 and Ph.D. in Electrical and Computer Engineering from the University of Manitoba, Canada in 2008. His research interests are in the areas of Internet of Things (IoT), machine learning, and incentive mechanism design.
\end{IEEEbiography}

\begin{IEEEbiography}[{\includegraphics[width=1in,height=1.25in,clip,keepaspectratio]{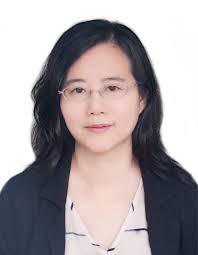}}]{Chunyan Miao} received the BS degree from Shandong University, Jinan, China, in 1988, and the MS and PhD degrees from Nanyang Technological University, Singapore, in 1998 and 2003, respectively. She is currently a professor in the School of Computer Science and Engineering, Nanyang Technological University (NTU), and the director of the Joint NTU-UBC Research Centre of Excellence in Active Living for the Elderly (LILY). Her research focus on infusing intelligent agents into interactive new media (virtual, mixed, mobile, and pervasive media) to create novel experiences and dimensions in game design, interactive narrative, and other real world agent systems.
\end{IEEEbiography}

\begin{IEEEbiography}[{\includegraphics[width=1in,height=1.25in,clip,keepaspectratio]{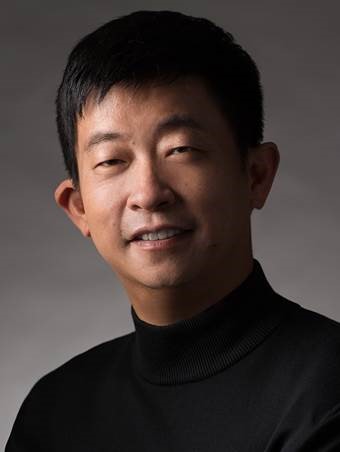}}]{Zhu Han} (S’01–M’04-SM’09-F’14) received the B.S. degree in electronic engineering from Tsinghua University, in 1997, and the M.S. and Ph.D. degrees in electrical and computer engineering from the University of Maryland, College Park, in 1999 and 2003, respectively. 

From 2000 to 2002, he was an R\&D Engineer of JDSU, Germantown, Maryland. From 2003 to 2006, he was a Research Associate at the University of Maryland. From 2006 to 2008, he was an assistant professor at Boise State University, Idaho. Currently, he is a John and Rebecca Moores Professor in the Electrical and Computer Engineering Department as well as in the Computer Science Department at the University of Houston, Texas. His research interests include wireless resource allocation and management, wireless communications and networking, game theory, big data analysis, security, and smart grid.  Dr. Han received an NSF Career Award in 2010, the Fred W. Ellersick Prize of the IEEE Communication Society in 2011, the EURASIP Best Paper Award for the Journal on Advances in Signal Processing in 2015, IEEE Leonard G. Abraham Prize in the field of Communications Systems (best paper award in IEEE JSAC) in 2016, and several best paper awards in IEEE conferences. Dr. Han was an IEEE Communications Society Distinguished Lecturer from 2015-2018, AAAS fellow since 2019, and ACM distinguished Member since 2019. Dr. Han is a 1\% highly cited researcher since 2017 according to Web of Science. Dr. Han is also the winner of the 2021 IEEE Kiyo Tomiyasu Award, for outstanding early to mid-career contributions to technologies holding the promise of innovative applications, with the following citation: ``for contributions to game theory and distributed management of autonomous communication networks."
\end{IEEEbiography}

\begin{IEEEbiography}[{\includegraphics[width=1in,height=1.25in,clip,keepaspectratio]{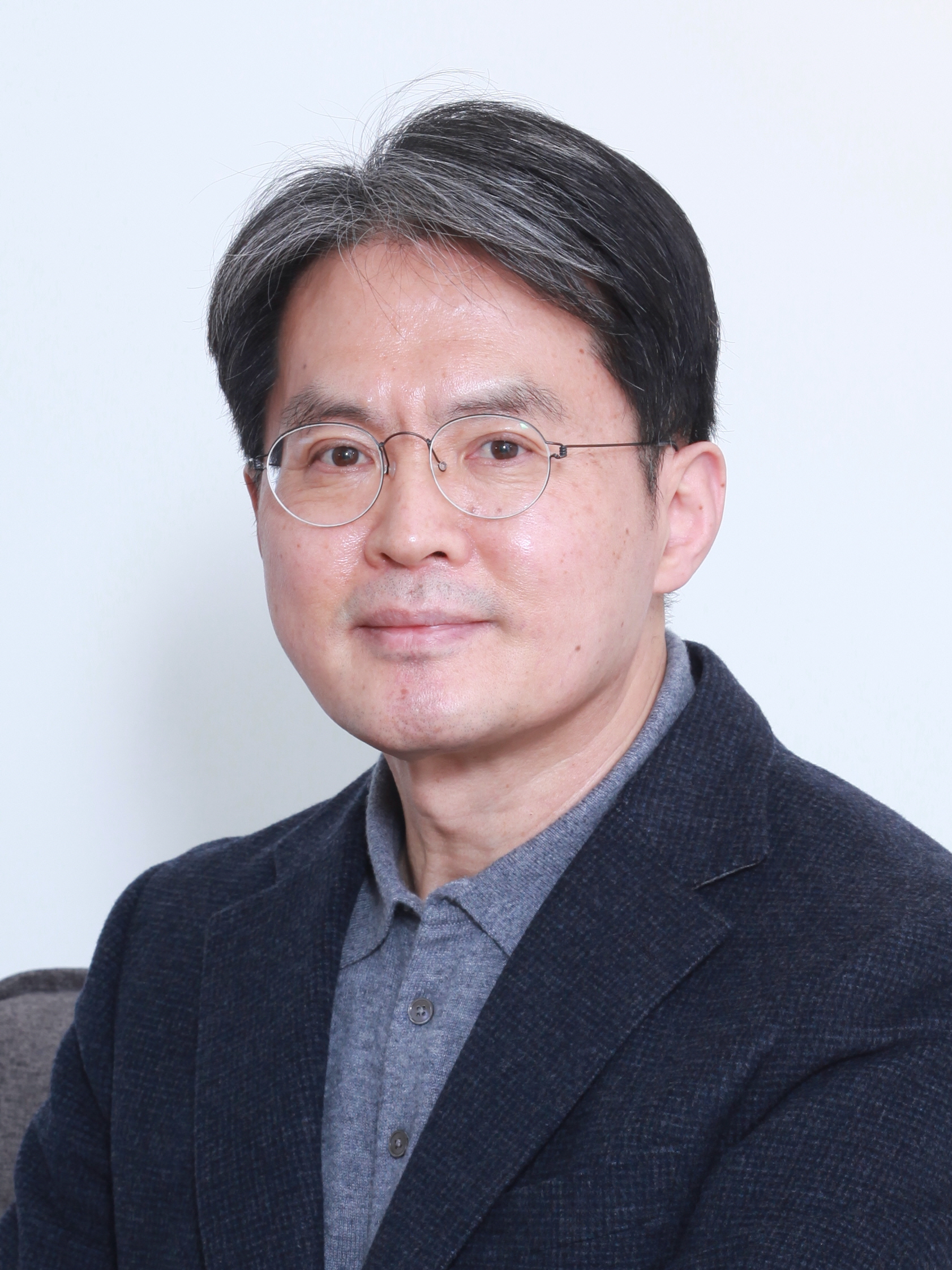}}]{Dong In Kim} (Fellow, IEEE) received the Ph.D. degree in electrical engineering from the University of Southern California, Los Angeles, CA, USA, in 1990. He was a Tenured Professor with the School of Engineering Science, Simon Fraser University, Burnaby, BC, Canada. Since 2007, he has been an SKKU-Fellowship Professor with the College of Information and Communication Engineering, Sungkyunkwan University (SKKU), Suwon, South Korea. He is a Fellow of the Korean Academy of Science and Technology and a Member of the National Academy of Engineering of Korea. He has been a first recipient of the NRF of Korea Engineering Research Center in Wireless Communications for RF Energy Harvesting since 2014. He has been listed as a 2020 Highly Cited Researcher by Clarivate Analytics. From 2001 to 2020, he served as an editor and an editor at large of Wireless Communications I for the IEEE Transactions on Communications. From 2002 to 2011, he also served as an editor and a Founding Area Editor of Cross-Layer Design and Optimization for the IEEE Transactions on Wireless Communications. From 2008 to 2011, he served as the Co-Editor-in-Chief for the IEEE/KICS Journal of Communications and Networks. He served as the Founding Editor-in-Chief for the IEEE Wireless Communications Letters, from 2012 to 2015. He was selected the 2019 recipient of the IEEE Communications Society Joseph LoCicero Award for Exemplary Service to Publications. He is the General Chair for IEEE ICC 2022 in Seoul.
\end{IEEEbiography}
\end{document}